# GeoSPM: Geostatistical parametric mapping for medicine

Holger Engleitner[1], Ashwani Jha[1], Marta Suarez Pinilla[1], Amy Nelson[1], Daniel Herron[2], Geraint Rees[1,3], Karl Friston[1], Martin Rossor[1], & Parashkev Nachev[1]

[1]UCL Queen Square Institute of Neurology, University College London, London, UK.

[2]Research & Development, NIHR University College London Hospitals Biomedical Research Centre, London, UK.

[3]Faculty of Life Sciences, University College London, Gower Street, London, UK.

## Abstract

The characteristics and determinants of health and disease are often organised in space, reflecting our spatially extended nature. Understanding the influence of such factors requires models capable of capturing spatial relations. Though a mature discipline, spatial analysis is comparatively rare in medicine, arguably a consequence of the complexity of the domain and the inclemency of the data regimes that govern it. Drawing on statistical parametric mapping, a framework for topological inference well-established in the realm of neuroimaging, we propose and validate a novel approach to the spatial analysis of diverse clinical data—GeoSPM—based on differential geometry and random field theory. We evaluate GeoSPM across an extensive array of synthetic simulations encompassing diverse spatial relationships, sampling, and corruption by noise, and demonstrate its application on large-scale data from UK Biobank. GeoSPM is transparently interpretable, can be implemented with ease by non-specialists, enables flexible modelling of complex spatial relations, exhibits robustness to noise and under-sampling, offers well-founded criteria of statistical significance, and is through computational efficiency readily scalable to large datasets. We provide a complete, open-source software implementation of GeoSPM, and suggest that its adoption could catalyse the wider use of spatial analysis across the many aspects of medicine that urgently demand it.

## 1 Introduction

Human beings vary along a rich multiplicity of social and biological dimensions, whose complex interactions across health and disease present a challenge for medical science and systems biology in general. The combination of large-scale data with machine learning promises to cast brighter light on this complexity than conventional inferential techniques, illuminating distributed, long-range dependencies hitherto obscured. Our interventions are increasingly grounded in an understanding of the factors that shape trajectories of disease evolution and determine individual responses to treatment.

One comparatively neglected dimension is the literal dimension of space: each of us inhabits a particular location that reflects and modifies our individual biological characteristics and the influence of (and on) other spatially distributed variables. Spatial factors may be static or vary over time, arising at multiple scales, ranging from the domestic to the inter-continental. Their reference frames may be set by internal communities, by external geographies, or by a complex blend of the two. Their spatial organisation may be linear or consistently distorted by individual or environmental movement within these frames of reference.

Spatial factors may disclose or alter characteristics of biology directly, or render them more or less clinically accessible or actionable. Space arises not only in epidemiology, environmental medicine, healthcare policy, and public health, but in the fundamental organisation of biology itself.

Yet outside a few specialist areas spatial analysis is comparatively rare in medicine. A comprehensive survey of Microsoft Academic Graph, spanning 30 years of medical research, reveals only 1897 journal papers at the intersection of spatial analysis and medicine, with an annual citation distribution for those cited more than once nonetheless substantially higher than a matched biomedical sample (mean 2.75 vs 2.13, Mann-Whitney U test, p<0.001, Figure S1). The scarcity is in part explained by the difficulty of the task. The spatial factors arising in a medical context are often entangled, their sampling is sparse and frequently corrupted by noise, and the underlying signals tend to be weak. But spatial analysis is hard even where the data regime is benign, for the problem is essentially multidimensional and is rarely, if ever, open to analytic solutions.

The fundamental challenge is reflected in the wide array of techniques in current use. A survey of 397 papers published since Jan 1$^{st}$ 2017 in the joint domains of health and spatial modelling identifies local indicators of spatial association (Anselin, 1995), spatial scan statistics (Kulldorff, 1997), inverse distance weighting (Shepard, 1968), kernel density estimation (Rosenblatt, 1956; Wand and Jones, 1994), spatial regression in terms of spatial lag and spatial error models (Anselin, 1988), geographically weighted regression (Brunsdon, Fotheringham and Charlton, 1998), land-use regression (Briggs *et al.*, 1997), kriging (Goovaerts, 1997; Stein, 1999), generalized linear mixed models (Breslow and Clayton, 1993), generalized (geo-)additive models (Hastie and Tibshirani, 1986; Kammann and Wand, 2003), hierarchical Bayesian spatial analysis (Banerjee *et al.*, 2003; Lawson, 2013), and model-based geostatistics (Diggle, Tawn and Moyeed, 1998; Diggle and Ribeiro, 2007), amongst others.

This methodological diversity reflects differing demands on the spatial aspects of the model. The most general and taxing research questions require methods that treat space as a continuity, produce spatially continuous estimates, and provide principled measures of uncertainty. Dominant in this category are methods that adopt a nonlinear multivariate approach, taking advantage of the flexibility and expressivity it offers. Though potentially powerful, they require joint expertise in the method and the domain of its application, depend on prior specification of model parameters, and tend to demand substantial computational resource even for data of moderate scale. Furthermore, in the generalized linear framework, space commonly enters the model as a latent random effect—usually derived from a suitable Gaussian process. This approach adjusts for spatially correlated variance within an otherwise non-spatial framework, with the fixed effects remaining constant across the spatial field (Gelfand *et al.*, 2003).

These obstacles motivate the pursuit of alternatives outside the multivariate paradigm. The direct counterpoint is a mass-univariate approach, where a complex multivariate model is replaced by a spatially indexed ensemble of simpler models. Geographically weighted regression (GWR) modifies the predictors in a regression model through a spatially localised weight matrix, so that a different model is estimated at each location. Although GWR estimates can be derived from a prespecified grid, in practice only sampled locations or grids of modest size tend to be evaluated owing to the difficulty of correcting for multiple comparisons in a topologically informed manner (da Silva and Fotheringham, 2016). Spatial inference with GWR is commonly limited to regression coefficient or coefficient of determination maps that simply indicate the local goodness of fit (Yuan *et al.*, 2020), without employing formal tests of significance (Brunsdon, Fotheringham and Charlton, 1996; Stewart Fotheringham, Charlton and Brunsdon, 1996). Finally, these are regression models relating a response to a set of spatially organised predictors, not models of the spatial variation of a set of variables within a topological framework of uncertainty.

Here we propose, implement, and validate a novel approach to the spatial analysis of diverse clinical data that draws upon differential geometry and random field theory. In particular, we leverage the procedures used in statistical parametric mapping (SPM): a framework for making *topological inferences* about spatially structured effects, with well-behaved spatial dependencies (Friston *et al.*, 1994). This approach has been established for decades in the realm of (structural and functional) volumetric neuroimaging.

The core idea is to transform sparse spatial signals into a form suited to mass-univariate statistical testing on a chosen point grid: for example, testing that the spatially or regionally expression of a particular variable is greater than would be expected, under the null hypothesis of no regional effect. The probability of observing topological features in the observed map, such as peaks or clusters (i.e., level sets above some threshold), can then be evaluated with classical inference based on random field theory, and used to ascribe a p-value to spatially organised effects. This principled approach radically simplifies spatial analysis, rendering it potentially more sensitive and robust to noise, and places it on a formal inferential footing, yielding a general-purpose geostatistical tool readily deployable across a multitude of medical fields.

In what follows, we (i) offer a detailed rationale for our approach; (ii) proceed to evaluate it across an extensive array of synthetic simulations where the nature of the spatial relationships, sampling, and corruption by noise are prespecified; and (iii) demonstrate its application on large-scale data from UK Biobank (https://www.ukbiobank.ac.uk/)(Sudlow *et al.*, 2015). The numerical analyses serve to establish face validity; the empirical analysis to demonstrate predictive validity. We provide a complete, open-source software implementation of our framework, released as an extension to SPM; namely, geospatial SPM or 'GeoSPM'.

## 2 Methods

### 2.1 Overview

Our approach builds on the well-established regression analysis framework implemented in SPM12 (http://www.fil.ion.ucl.ac.uk/spm/), the most widely used platform for spatial inference in brain imaging. Within this framework, a set of explanatory variables is associated with a multivariate, spatially-structured response, whose components represent measurements taken at regular locations in a spatial domain. The association between explanatory variables and response is estimated at each location separately, using the same general linear model (GLM). This yields a collection of univariate multiple regression models that share the same design matrix but differ in the model parameters used to explain the response or dependent variable. Crucially, random fluctuation, or variations in the response variable that are not explained by the GLM, are treated as realisations of a random (spatial) field with certain contiguity or smoothness properties. This is *mass-univariate* inference from a spatial perspective.

A distinguishing feature of SPM is the manner of correcting for multiple comparisons when testing mass-univariate model parameters (i.e., regression coefficients) for significance. The large number of tests, performed simultaneously, gives rise to a proportionally large number of false positives by chance alone. Conversely, the strong spatial correlations among the components of the response violate assumptions of mutual independence, and render simple Bonferroni corrections inappropriately strict. SPM applies a more suitable correction by modelling the residuals as a random Gaussian field, so that p-values are meaningful in terms of identifying significant peaks and clusters in a discretised spatial domain. Heuristically, topological inference of this kind automatically accounts for spatial dependencies; in the sense that smooth random fluctuations will produce a smaller number of maxima than rough random fields with less spatial dependence (even though the total area above some threshold could be the same).

The kind of data we are concerned with comprises variables of interest observed at locations in a continuous spatial domain $D$. $D$ is usually a subset of $\mathbb{R}^2$ representing coordinates of a geographic space. More precisely, every element in a spatially-referenced data set associates a vector $\boldsymbol{y}_i$ of $P$ variable observations $(y_{i1}, \ldots, y_{iP})^{\mathrm{T}} \in \mathbb{R}^P$ with a location $\boldsymbol{x}_i \in D$

$$(\boldsymbol{y}_i, \boldsymbol{x}_i): i = 1, \ldots, N.$$

SPM typically requires data sampled at regular locations across a grid, spanning the spatial domain. However, we wish to analyse data that are irregularly and sometimes sparsely sampled. This can be resolved by distributing each data point locally—over regular grid locations—using a spatial Gaussian kernel of suitable and fixed variance.

From a data-centric point of view, we can interpret this spatial transformation as estimating the contribution of an individual observation to regular sample points, where the contribution has a maximum value at the observation location and then diminishes with increasing distance. In this way, the dependent variable in the univariate regression at any location of space is essentially a weighting of individual observations according to their proximity to that location: the higher the local response, the closer the observation. We can do this with impunity because we are interested in the explainable differences in these contributions at prespecified (grid point) locations. These explainable differences are assessed with normalised effect sizes (i.e., classical statistics), which are not affected by the total contribution or variance (Friston *et al.*, 1994, 2000)

The chosen variance of the Gaussian kernel is a parameter—hereafter called the *smoothing* parameter—deliberately left open to the analyst to specify the appropriate degree of spatial coarse graining (i.e., spatial smoothness of the data features in question). Since SPM naturally handles volumetric data, we are free to select multiple smoothing values on a continuous positive scale, rendering them as different spatial 'scales' or 'features' of a response variable (Worsley *et al.*, 1996). Here, two coordinates represent the location in space (i.e., location space), and the third coordinate tracks spatial spread (i.e., scale space), allowing the regression analysis to operate at different scales simultaneously. It is appropriate to permit inference under varied assumptions of uncertainty, allowing the analyst to draw conclusions from the similarities and differences obtained across the range of plausible spatial scales. The analyst is also free to implement mechanisms that select an optimal parameter under some criterion: here we suggest one pragmatic method of doing this. Note that this scale-space implementation of topological inference automatically accounts for dependencies in moving from one scale to another and enables topological inference in terms of maxima or clusters in both location and scale space (i.e., a particular effect can be declared significant at this location and this spatial scale). For simplicity, we will focus on topological inference at a given spatial scale.

Downstream of the above spatial transformation of data features, the statistical approach is formally identical to a standard SPM analysis. The output comprises a series of volumes representing regression coefficients, statistical contrasts derived from these model parameters, the statistical parametric maps—of classical statistics based on these contrasts—and, lastly, thresholded binary maps that indicate whether the voxels in the corresponding statistical map are significant at the chosen (suitably corrected) *p*-value.

**2.2 Synthetic data and generative models**

The statistical validity of the proposed approach is underwritten by the assumptions on which SPM rests. Nonetheless, it is helpful to examine its construct validity, in comparison with alternative methods (e.g., kriging), and face validity, in terms of its ability to recover known effects in different situations. Such validation is best performed with a known (spatial) ground truth, under manipulations of sampling and noise wide enough to cover typical use cases.

Here, we use synthetic data drawn from a generative model with a spatially varying distribution of one or joint two binary variables. The spatial variability of the distribution is determined by fractal shapes. Fractals characteristically exhibit detail across an infinite range of spatial scales, which makes them suitable candidates for a spatially-structured ground truth. The use of binary response variables allows us to focus on data that are generated in a spatially structured way; namely, in a regionally specific fashion under various levels of noise or stochasticity.

We start by defining a spatial domain as a rectangular subset $D := [0,a) \times [0,b) \subset \mathbb{R}^2$ for some $a, b \in \mathbb{N}^+$, and restrict $a$ and $b$ to positive integers so that

$$D = [0,a) \times [0,b) = \bigcup_{j=1}^{a} \bigcup_{k=1}^{b} [j-1,j) \times [k-1,k)$$

has a natural decomposition into $a \times b$ grid cells with coordinates $(j,k) \in D' := \{1, \ldots, a\} \times \{1, \ldots, b\}$. Assuming that there are $P$ binary factors in the generative model, the simulated response variables can be written as the components of a random vector $\mathbf{Z} = (Z_1, \ldots, Z_P)^\mathrm{T} \in \{0,1\}^P$, which is sampled at random grid cells $\mathbf{W} \in D'$ in the underlying space. Our data generation mechanism is based on the factorisation of the joint distribution of $\mathbf{Z}$ and $\mathbf{W}$ as

$\Pr(\mathbf{Z}, \mathbf{W}; \boldsymbol{\theta}) = \Pr(\mathbf{Z} \mid \mathbf{W}; \boldsymbol{\theta}) \Pr(\mathbf{W})$,

so that the response variables $(Z_1, \ldots, Z_P)^\mathrm{T}$ are conditioned on location $\mathbf{W}$ with model parameters $\boldsymbol{\theta}$ fixed at $(\theta_1, \theta_2, \ldots)^\mathrm{T}$. $\mathbf{W}$ is distributed independently of $\mathbf{Z}$ and uniformly over $D'$.

The conditioning allows breaking down the distribution of $\mathbf{Z}$ spatially, by partitioning the grid $D'$ into a small number of (not necessarily) continuous regions $R_k \subseteq D' : k = 0, \ldots, K-1$ for which local distributions $\mathrm{P}_k(\mathbf{Z}; \boldsymbol{\theta}) := \Pr(\mathbf{Z} \mid \mathbf{W} \in R_k; \boldsymbol{\theta})$ can be specified for each region, $k$. As $P \in \{1,2\}$ for the models considered here, at most four probabilities are required for each of these local distributions. The partitions are based on arrangements of fractal shapes, shown in Figure 1 for the two bivariate models and in Figure 2 for the univariate models.

In these images, a single pixel represents a grid cell, and the shading indicates the distinct distributions $\mathrm{P}_k(\mathbf{Z}; \boldsymbol{\theta})$. The resolution of the bivariate models is 220 by 210 grid cells, whereas the resolution of the univariate models is 120 by 120 grid cells. Geometry for the fractal shapes is constructed by recursively substituting the edges of a (start) shape with a simple curve (Figure 3) and then rasterising the resulting polygon into the grid using MATLAB's poly2mask function. We wish to examine the effect of noise and interactions in our numerical experiments, which leads us to consider two distinct parameterisations of the distributions $P_k$.

### 2.2.1 Parameterisation of $P_k(Z;\theta)$ for Examining Noise

The first parameterisation is expressed in terms of a function $p_{noise}(\cdot)$ with parameters $p$ and $q$:

$$P_k(Z;\theta) = p_{noise}(Z_1, Z_2; p, q), \quad p = \theta_k, q = \theta_{k+K}$$

$p_{noise}(\cdot)$ is summarised in Table 1. The parameters $p$ and $q$ are simply the values of the marginal probabilities $p_{noise}(z_1 = 1)$ and $p_{noise}(z_2 = 1)$ and are sufficient for defining $p_{noise}(Z_1, Z_2; p, q)$ if $Z_1$ and $Z_2$ are assumed to be independent.

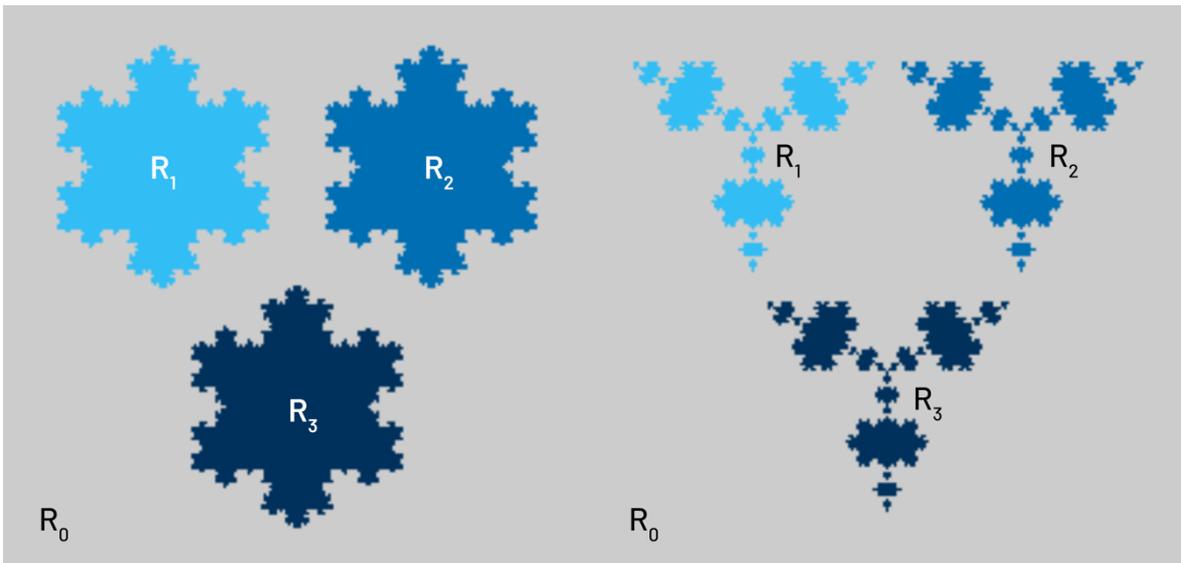

**Figure 1.** The four distinct regions $R_k : k = 0, ..., 3$, of the joint conditional probability $\Pr(Z | W; \theta)$ for the snowflake model (left) and the anti-snowflake model (right).

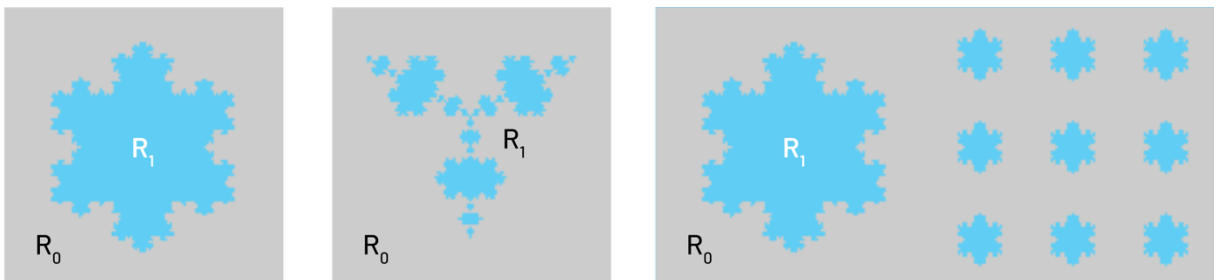

**Figure 2.** The two distinct regions $R_k : k = 0, 1$ of the conditional probability $\Pr(Z | W; \theta)$ for the univariate models: snowflake model (left), anti-snowflake model (middle) and snowflake field model (right).

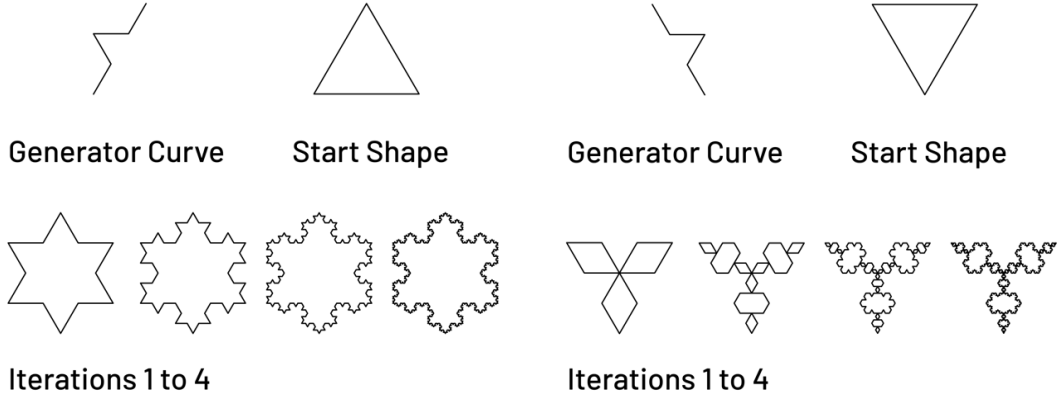

**Figure 3.** The geometric construction used for the fractal shapes. Koch snowflake on the left, Koch anti-snowflake on the right.

| $p_{noise}(Z_1, Z_2; p, q)$ | $z_1 = 0$ | $z_1 = 1$ | $p_{noise}(z_2)$ |
|---|---|---|---|
| $z_2 = 0$ | $(1-q)(1-p)$ | $(1-q)p$ | $1-q$ |
| $z_2 = 1$ | $q(1-p)$ | $qp$ | $q$ |
| $p_{noise}(z_1)$ | $1-p$ | $p$ | $1$ |

**Table 1.** Probability table for the bivariate local distribution function $p_{noise}(Z_1, Z_2; p, q)$

As we move in $D$ from one region to the next, we simulate spatially distinct conditions of the variables $Z_1$ and $Z_2$ by changing the regional expectations $\mathbb{E}_k[(Z_1, Z_2)]$ through $p_{noise}(Z_1, Z_2; p, q)$: For the four regions of the bivariate models in Figure 1, the corresponding expectations are listed in Table 2, together with the respective values for $p$ and $q$. In the absence of uncertainty, the models generate the expected values in each region exactly, thus $R_0$ *only* generates observations $(0, 0)$, $R_1$ *only* $(1, 0)$, and so on.

As more uncertainty (i.e., noise) is introduced—by adjusting the values for $p$ and $q$—the overall pattern of observations still holds, but other values have a non-zero probability of occurrence: $R_0$ *mostly* generates observations $(0, 0)$, $R_1$ *mostly* $(1, 0)$, and so on, until a maximum level of uncertainty is reached and each observation is equally probable in every region. By expressing $p$ and $q$ in terms of a single parameter $\gamma \in [0, \ldots, 0.5]$ (the last two columns in Table 2), we can easily vary the degree of observation noise from a spatially deterministic and regionally differentiated form, to one where all regional differentiation is lost (see Figure 4 for an example).

| | $\mathbb{E}_k[(Z_1, Z_2)] \rightarrow \ldots$ | $p \rightarrow \ldots$ | $q \rightarrow \ldots$ | $p(\gamma)$ | $q(\gamma)$ |
|---|---|---|---|---|---|
| $R_0$ | $(0,0)$ | $0$ | $0$ | $\gamma$ | $\gamma$ |
| $R_1$ | $(1,0)$ | $1$ | $0$ | $1-\gamma$ | $\gamma$ |
| $R_2$ | $(0,1)$ | $0$ | $1$ | $\gamma$ | $1-\gamma$ |
| $R_3$ | $(1,1)$ | $1$ | $1$ | $1-\gamma$ | $1-\gamma$ |
| $max. uncertainty$ | $(0.5, 0.5)$ | $0.5$ | $0.5$ | | |

**Table 2.** Expected values of $Z_1$ and $Z_2$ in each region of the Snowflake and Anti-Snowflake models shown in Figure 1 for the given values of $p$ and $q$.

The parameter vector $\boldsymbol{\theta}$ in $P_k(Z; \boldsymbol{\theta})$ for the bivariate models is determined by $\gamma$ as shown in Table 2 and has the following structure:

$$\boldsymbol{\theta} = \boldsymbol{\theta}_\gamma, \qquad \boldsymbol{\theta}_\gamma := [\gamma, 1-\gamma, \gamma, 1-\gamma, \gamma, \gamma, 1-\gamma, 1-\gamma]$$

We consider the deviation of observed values of $Z$ when $\gamma$ is non-zero from the expected values when $\gamma$ is 0 as simulating noise induced by confounding variables that are not captured in the data. Its effect of degrading the observable spatial differentiation of the variables of interest is key in our analysis of the performance of GeoSPM, and so we treat $\gamma$ as an independent variable in these numerical experiments.

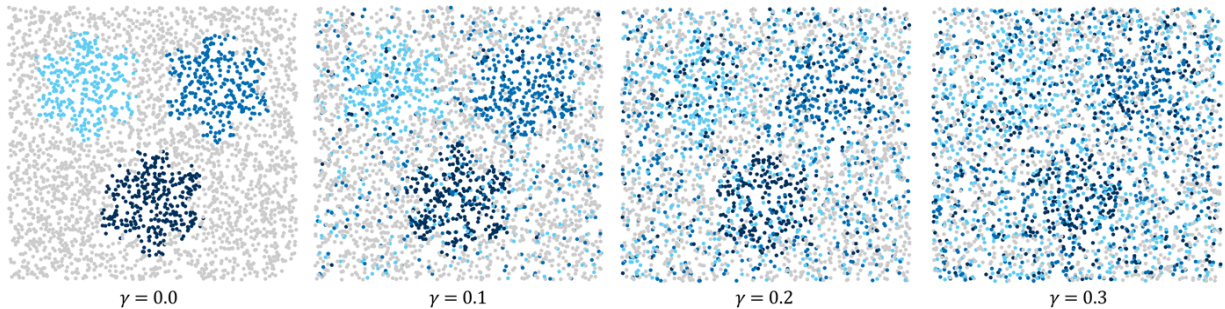

**Figure 4**. Random realisations of the bivariate Snowflake model for different levels of $\gamma$ (from left to right).

Of course, real data also exhibit additive measurement noise. To simulate measurement noise an observation $Y \in \mathbb{R}^P$ with location $X \in D$ is derived from $Z$ and $W$ by adding random effects sampled from multidimensional uniform distributions:

$$Y = Z + \zeta, \qquad \zeta \sim \text{uniformly on } I_1 \times \ldots \times I_P, \qquad I_i = [0, 0.005]$$
$$X = W + \omega, \qquad \omega \sim \text{uniformly on } [0,1) \times [0,1)$$

A practical benefit of applying 'spatial noise' $\omega$ to $W$ is that the probability of two randomly drawn elements $(y_i, x_i)$ and $(y_j, x_j)$ coinciding at the same location $x_i = x_j$ is minimised. This is relevant for the geostatistical method used for validation in these numerical experiments, because such collisions would produce singular, non-positive definite covariance matrices, when predicting observations and need to be removed from any data set prior to model estimation.

As GeoSPM only operates in terms of the discrete space $D'$, it always applies the congruency $X \equiv W$ and so the added noise $\omega$ has no effect. We will also consider an alternative method for resolving spatially coincident observations—by averaging observations from the same location—and will provide corresponding results for kriging in supplementary materials.

We can now summarise the procedure for generating a spatially-referenced data set of size $N$ and noise level $\gamma_0$ as follows:

1. Draw a uniform sample of $N$ grid cell coordinates $w_i \in \{1, \ldots, a\} \times \{1, \ldots, b\}$, $i: 1, \ldots, N$
2. For each grid cell coordinate $w_i$ draw a sample $z_i$ from $P_k(Z; \theta_\gamma)$, where $w_i \in R_k$
3. Obtain an observation $y_i$ at location $x_i$ by adding small amounts of random noise to $z_i$ and $w_i$:
$$y_i = z_i + \zeta_i$$
$$x_i = w_i + \omega_i$$

### 2.2.2 Parameterisation of $P_k(Z; \theta)$ for Examining Interactions

A common feature of regression modelling is the inclusion of interaction terms, when there is reasonable belief that the marginal effect of one variable depends on the value of another. In a spatial setting, an interaction could be described as the degree to which the observation of a value of one variable is affected by the value of another variable at the same location. Therefore, GeoSPM's ability to detect interactions merits additional evaluation.

Here, a second parameterisation of the local distributions $P_k$ can be motivated by interpreting the spatial response introduced earlier as a *concentration* instead of a measure of closeness. The constituent probabilities of the $P_k$ are then the *regionally expected concentrations* of their respective observations $(Z_1, Z_2) \in \{(0,0), (1,0), (0,1), (1,1)\}$. As the response of local univariate regression models, these concentrations should have approximately *additive structure*, given the objective is to model interactions. To this end, we define the $P_k$ as a function $p_{interaction}(\cdot)$ with parameters $p_0, c_1, c_2$ and $c_3$, which we assign from a global parameter vector $\boldsymbol{\theta}$ for each region $R_k$:

$$P_k(\mathbf{Z}; \boldsymbol{\theta}) = p_{interaction}(Z_1, Z_2; p_0, c_1, c_2, c_3), \qquad p_0 = \theta_k, c_i = \theta_{k+iK}$$

Table 3 provides a definition of $p_{interaction}(\cdot)$. The parameters $c_i$ approximate the *effect sizes* induced by the observations of the variables $(Z_1, Z_2)$ in the local univariate regression models, which—due to their binary values—correspond to the effect of $Z_1$, $Z_2$ and their interaction $Z_1 \times Z_2$ on the concentration or response. This is only an approximation of effect size, because of the non-linear fall-off of the Gaussian kernels used to synthesize the response.

| $p_{interaction}(Z_1, Z_2; p_0, c_1, c_2, c_3)$ | $z_1 = 0$ | $z_1 = 1$ |
|---|---|---|
| $z_2 = 0$ | $p_0$ | $p_0 + c_1$ |
| $z_2 = 1$ | $p_0 + c_2$ | $p_0 + c_1 + c_2 + c_3$ |

**Table 3.** Probability table for the bivariate local distribution function $p_{interaction}(Z_1, Z_2; p_0, c_1, c_2, c_3)$

The parameters $c_i$ define the respective probabilities relative to the parameter $p_0$, the probability of generating the observation $(0, 0)$. The $c_i$ have a maximum range of $[-1, ..., 1]$, subject to the conditions that each $p_i$ is a valid probability ($p_i \in [0, ..., 1]$) and that all the resulting probabilities add up to 1:

$$p_1 = p_0 + c_1$$

$$p_2 = p_0 + c_2$$

$$p_3 = p_0 + c_1 + c_2 + c_3$$

$$1 = 4p_0 + 2c_1 + 2c_2 + c_3$$

For the interaction experiments, we define the same four distinct regions $R_{0...3}$ as for the bivariate snowflake model (shown on the left of Figure 1) in the noise parameterisation and vary the magnitude of the interaction effect $c_3$ in region $R_3$. In detail, in region $R_0$, all observations are equiprobable, representing the null state, as all three effects $c_1, c_2$ and $c_3$ are 0. $R_1$ and $R_2$ are regions where we either observe a non-zero effect $c_1$ for variable $Z_1$ or a non-zero effect $c_2$ for variable $Z_2$ but no effect that corresponds to their interaction.

Finally, we construct region $R_3$ to model an interaction effect $c_3$ at different intensities while keeping $p_0$ constant at a non-zero level and assuming $c_1 = c_2$. Based on (arbitrarily) setting $p_0$ to a small non-zero value of 0.025 and obeying all constraints, $c_3$ can range between 0 and 0.9. Given $p_0$ and a value for $c_3$ we can derive $c_1 = c_2 = \frac{1 - 4p_0 - c_3}{4}$. The regional probabilities chosen for the experiments are summarised in Table 4 (including null values for when $c_3 = 0$), where we picked 6 equally spaced settings between 0.25 and 0.5 for the interaction effect $c_3$ in $R_3$.

|       | $R_0$ | $R_1$ | $R_2$ | $R_{3[null]}$ | $R_{3[0.25]}$ | $R_{3[0.3]}$ | $R_{3[0.35]}$ | $R_{3[0.4]}$ | $R_{3[0.45]}$ | $R_{3[0.5]}$ |
|---|---|---|---|---|---|---|---|---|---|---|
| $p_0$ | 0.25 | 0.125 | 0.125 | 0.025 | 0.025 | 0.025 | 0.025 | 0.025 | 0.025 | 0.025 |
| $c_1$ | 0 | 0.25 | 0 | 0.225 | 0.1625 | 0.15 | 0.1375 | 0.125 | 0.1125 | 0.1 |
| $c_2$ | 0 | 0 | 0.25 | 0.225 | 0.1625 | 0.15 | 0.1375 | 0.125 | 0.1125 | 0.1 |
| $c_3$ | 0 | 1 | 1 | 0 | 0.25 | 0.3 | 0.35 | 0.4 | 0.45 | 0.5 |
| $p_1$ | 0.25 | 0.375 | 0.125 | 0.25 | 0.1875 | 0.175 | 0.1625 | 0.15 | 0.1375 | 0.125 |
| $p_2$ | 0.25 | 0.125 | 0.375 | 0.25 | 0.1875 | 0.175 | 0.1625 | 0.15 | 0.1375 | 0.125 |
| $p_3$ | 0.25 | 0.375 | 0.375 | 0.475 | 0.6 | 0.625 | 0.65 | 0.675 | 0.7 | 0.725 |

**Table 4.** Approximate effect sizes and derived probabilities for each region of the interaction experiments. Region $R_3$ is the only region that is varied across experiments, by increasing the magnitude of the interaction effect $c_3$ specified in square brackets.

As with the noise parameterisation, a data set of $N$ observations can be generated in a few simple steps. Again, we add random effects $\boldsymbol{\zeta}$ and $\boldsymbol{\omega}$ from multidimensional uniform distributions as defined above:

1. Draw a uniform sample of $N$ grid cell coordinates $\boldsymbol{w}_i \in \{1, \dots, a\} \times \{1, \dots, b\}, \; i: 1, \dots, N$
2. For each grid cell coordinate $\boldsymbol{w}_i$ draw a sample $\boldsymbol{z}_i$ from $\mathrm{P}_k(\boldsymbol{Z}; \boldsymbol{\theta}_{interaction})$, where $\boldsymbol{w}_i \in R_k$ and $\boldsymbol{\theta}_{interaction}$ is the vector of combined regional parameters $p_0, c_1, c_2$ and $c_3$ for all four regions.
3. Obtain an observation $\boldsymbol{y}_i$ at location $\boldsymbol{x}_i$ by applying small amounts of random noise to $\boldsymbol{z}_i$ and $\boldsymbol{w}_i$:

$$\boldsymbol{y}_i = \boldsymbol{z}_i + \boldsymbol{\zeta}_i$$
$$\boldsymbol{x}_i = \boldsymbol{w}_i + \boldsymbol{\omega}_i$$

This completes our description of the synthetic data used to establish the face validity of GeoSPM. We now return to the empirical data used to provide an illustrative application and provisional predictive validity.

**2.3 Demonstration with UK Biobank data**

To demonstrate the application of GeoSPM to real data, we chose to explore the potential association between a common disease type 2 diabetes, and a small number of demographic variables in UK Biobank drawn from the area of Greater Birmingham. It should be stressed that the sole purpose of this application was to illustrate its application. It is not our intention to make inferences about the data itself. The objective instead is to show how spatial variation of a variable of interest can be examined, while accounting for the effect of commonly associated factors. Any remaining spatial structure points to the presence of additional factors not yet considered or known.

UK Biobank provides a large collection of health and genetic information for its prospective cohort of more than 500 000 participants recruited between 2006 and 2010 with assessment centres throughout Great Britain (https://www.ukbiobank.ac.uk/) (Sudlow *et al.*, 2015).

We extracted a set of variables from UK Biobank in a region defined by a 35 km by 35 km square (spanning from 388000E, 423000N in the south-west corner to 269000E, 304000N in its north-east corner, in coordinates of the Ordnance Survey National Grid). The variables were sex (field 31), age (field 21022), body mass index (BMI, field 21001), household income (field 738) and the location of the participants (fields 20074 and 20075). Location information is based on the postcode for the address to which the participants invitation was sent. The location co-ordinates use the Ordnance Survey reference, rounded to the nearest kilometre. UK Biobank provides one or more temporal instances for certain fields. For such fields, the value of the earliest instance was chosen, which was the case for BMI and household income. In addition, ICD-10 and ICD-9 diagnosis codes were gathered from a separate hospital inpatient data table named HESIN_DIAG provided through field 41259. From these diagnosis codes we defined an indicator variable for type 2 diabetes, whose value was set to 1 whenever a participant had a record of either an ICD-10 code in block E11 ("type 2 diabetes mellitus") or at least one of a handful of relevant ICD-9 codes as specified in Table 7 in supplementary materials. The number of participants with available data for all selected variables in the selected area of Birmingham was 18193.

As a preliminary sanity check for the presence and degree of associativity, the diabetes indicator variable was entered as the response variable into a multiple Bayesian logistic regression model with a ridge prior. Sex, age, BMI, household income and the interaction between BMI and household functioned as predictors. Age, BMI and household income were centred at 0 and divided by their respective sample standard deviations. The interaction term was then formed as a simple multiplication. The model was evaluated by BayesReg version 1.9.1 (Makalic and Schmidt, 2016) in MATLAB. BayesReg uses a Markov Chain Monte Carlo (MCMC) Gibb's sampler. Posterior parameters were estimated from a single chain of 250000 samples (after a burn-in period of the same number of samples), of which only every 5$^{th}$ sample was used for computing the estimate. The posterior means of the regression coefficients and their credible intervals were as follows:

| Predictor | Coefficient Posterior Mean ± SD | 95% Credible Interval | t-Statistic | ESS |
|---|---|---|---|---|
| Sex | 0.748 ± 0.054 | (0.642 to 0.855) | 13.79 | 82.2 |
| Age | 0.313 ± 0.029 | (0.257 to 0.371) | 10.73 | 83.5 |
| BMI | 0.719 ± 0.025 | (0.670 to 0.769) | 28.64 | 71.5 |
| Household Income | -0.344 ± 0.036 | (-0.416 to -0.274) | -9.50 | 61.9 |
| BMI x Household Income | 0.053 ± 0.028 | (-0.001 to 0.108) | 1.93 | 73.0 |

**Table 5.** Results of the preliminary Bayesian logistic ridge regression analysis of the UK Biobank diabetes data set extracted for Birmingham.

The results showed that there is a reasonably strong association between type 2 diabetes and all main terms, but evidence for an interaction between BMI and household income appears to be weak. On the basis of this preliminary analysis, we directed our attention to the spatial variability of diabetes and the question of how much of this spatial variability is driven by the other variables. We defined a progression of four models, listed in Table 6.

| Model | Type 2 Diabetes | Sex | Age | BMI | Household Income | BMI x Household Income |
|---|---|---|---|---|---|---|
| 1 | ■ | – | – | – | – | – |
| 2 | ■ | ■ | ■ | ■ | – | – |
| 3 | ■ | ■ | ■ | ■ | ■ | – |
| 4 | ■ | ■ | ■ | ■ | ■ | ■ |

**Table 6.** The four GeoSPM models used for the Birmingham data from UK Biobank.

It is important to keep in mind that unlike in the preliminary analysis, in these GeoSPM models, type 2 diabetes is no longer a response variable but an explanatory or independent variable, which means its effect is marginalised relative to the other variables in each model. By applying a single colour map to all regression coefficient maps across models, the intensity and nature of topological changes—in the marginalised contribution of each variable—become visible, not only within a single model but over the ensemble of four models. Similarly, changes in the extent and location of significant areas, due to the addition of variables as we move from one model to the next, allow us to assess patterns of spatial variability. Lastly, using intersections between significant areas of multiple variables, we can identify areas of significant *conjunctions* between those variables (Friston *et al.*, 1999).

## 2.4 Numerical experiments

### 2.4.1 Kriging

We evaluate GeoSPM in comparison with the multivariate geostatistical method of kriging. Kriging (Goovaerts, 1997) is an ensemble of linear least-squares regression techniques for predicting the value of a random field at an unsampled location from observations at other locations. It is commonly used when interpolating spatially-referenced point data over a surface and provides a measure of the uncertainty in its predictions. In statistics and machine learning, kriging is essentially an application of multivariate Gaussian process prediction. Crucially, kriging requires an explicit model of the spatial covariance and cross-covariance of the data, which needs to be chosen *a priori*. As the random field is generally assumed to be second-order stationary and isotropic, the covariance can be expressed as a function of the Euclidean distance between a pair of points, independently of their actual location in the spatial domain. Parameters required by the selected covariance function are estimated from the data and substituted for the true values when computing the predictions. The covariance and cross-covariance model we used for all kriging predictions is the family of Matérn functions (Matérn, 1960) together with an added "nugget" component. The Matérn model exhibits adaptable smoothness controlled by a parameter $\kappa$ and is recommended as a sensible default choice in the literature (Diggle and Ribeiro, 2007; Stein, 1999). The nugget component adds a discontinuous jump to the covariance function at coincident points and captures variance due to measurement error. Both, the Matérn model and the nugget component have a sill parameter that determines the strength of their contribution to the covariance, while the former also incorporates a standard range parameter reflecting its spatial scale in addition to the parameter controlling smoothness.

All kriging computations were done in R using the gstat package (Pebesma, 2004), which is available at https://cran.r-project.org/web/packages/gstat/index.html. In cases where the experimental data contained several variables, gstat estimated a linear model of coregionalization (LCM), which expressed all required auto- and cross-covariances as linear combinations of a Matérn function and a nugget component. The range parameter is constrained by gstat to be the same for all covariances in the LCM and was estimated from the first variable in the data prior to estimating the LCM. We configured gstat to use ordinary (co-)kriging with a constant but unknown mean in a global search window.

For each variable of interest kriging produced an image of the predicted mean and an image of the corresponding prediction variance, which is derived solely from the arrangement of positions in the data.

### 2.4.2 Synthetic Experiments: Noise Parameterisation

The numerical face validation experiments are based on three univariate models (snowflake, anti-snowflake, snowflake field) and two bivariate models (snowflake, anti-snowflake) as depicted in Figure 1 and 2. For all models, we ran experiments at different sampling levels, $N_{univariate} \in \{600, 1200, 1800\}$ and $N_{bivariate} \in \{1600, 3200\}$ and increased the noise parameter $\gamma$ from 0.0 to 0.35 in 0.01 increments. For each triplet (model, $N$, $\gamma$), 10 independent data sets were randomly generated.

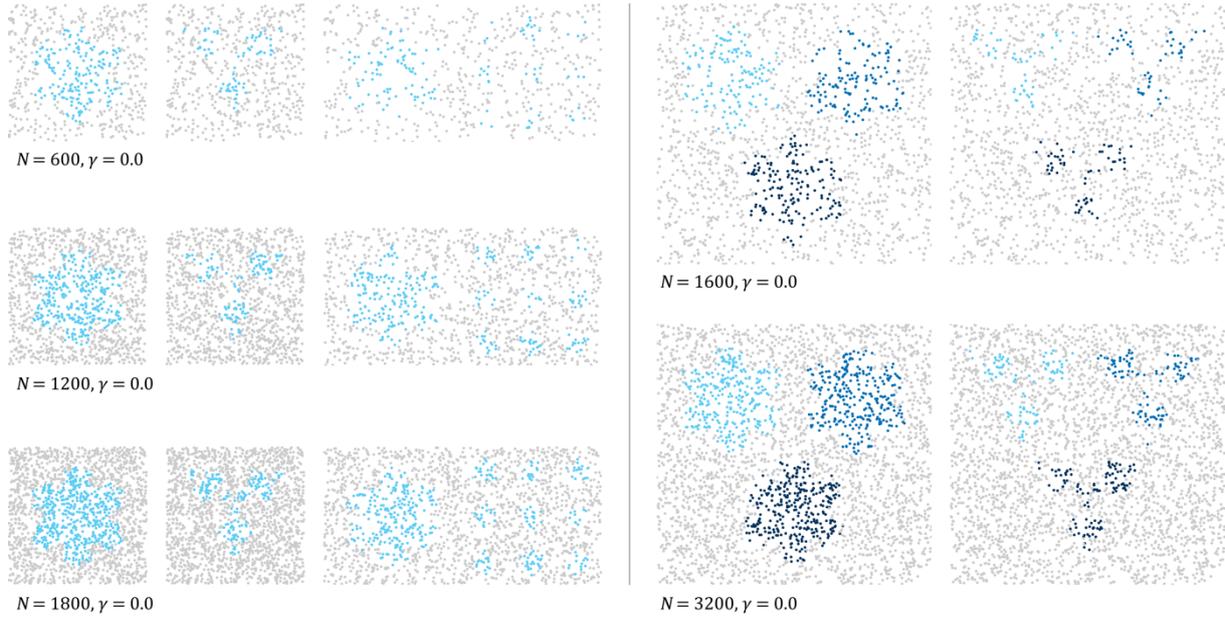

**Figure 5**. Sampling levels (noise-free, $\gamma = 0.0$) for the univariate models on the left ($N = 600, 1200, 1800$), and for the bivariate models on the right ($N = 1600, 3200$).

Each generated data set was processed by GeoSPM as well as gstat. For GeoSPM, the spread of the spatial response at locations $x_i$ was modelled at increasing smoothing parameter values ($\ell = 10$) using the 95 percent iso-density diameters of the bivariate normal distribution, $s = (10, 15, 20, 25, 30, 35, 40, 45, 50, 55, 60)^T$. The largest value of the smoothing parameter of 60 was chosen to be half the height of the grid for the univariate models. The regression coefficients estimated by GeoSPM were tested using a one-tailed $t$-test at $P < 0.05$ FWE (voxel-level, family-wise correction), producing a stack of $\ell$ binary maps of significant areas for every variable of interest. In order to derive corresponding maps – one per variable – for kriging, we compared a standardised form of the kriging prediction $\hat{y}_{std}(j, k)$ to the critical value of the upper tail probability $P < 0.05$ of the normal distribution. We standardised $\hat{y}(j, k)$ at each grid cell $(j, k)$ using its estimated (positional) variance $\hat{\sigma}(j, k)$ and assuming a null mean of 0.5 to produce $\hat{y}_{std}(j, k)$:

$$\hat{y}_{std}(j, k) = \frac{\hat{y}(j, k) - \mu_{null}}{\hat{\sigma}(j, k)}, where (j, k) \in D', \mu_{null} = 0.5.$$

For a fair comparison with kriging, one of the $\ell$ smoothing values and its associated maps produced in a run of GeoSPM had to be chosen. We based this choice on maximising the spatial coverage by the significant areas at each spatial scale (cf. Figure 6), while minimising the spatial overlap between them. A spatial condition in the context of the observed variables $Y \in \mathbb{R}^P$ in our models is obtained by applying a threshold of 0.5 to all observations, recording as 1 if an observed variable value exceeds the threshold or 0 if it does not. Each observation of a univariate model can thus be assigned one of two spatial conditions, or one of four conditions in the case of a bivariate model. We obtain the significant areas for each spatial condition by running a separate analysis in GeoSPM on a set of data that represents the spatial condition of each observation as a one-hot encoding.

This approach enabled us to derive a score for each of the $\ell$ smoothing values, which simply comprised the total number of significant grid cells that appeared for exactly one of the spatial conditions, thereby ignoring any overlap. The smoothing value with the highest score was selected, together with the binary maps of significant areas computed from it. Ties were broken by choosing the smallest scale.

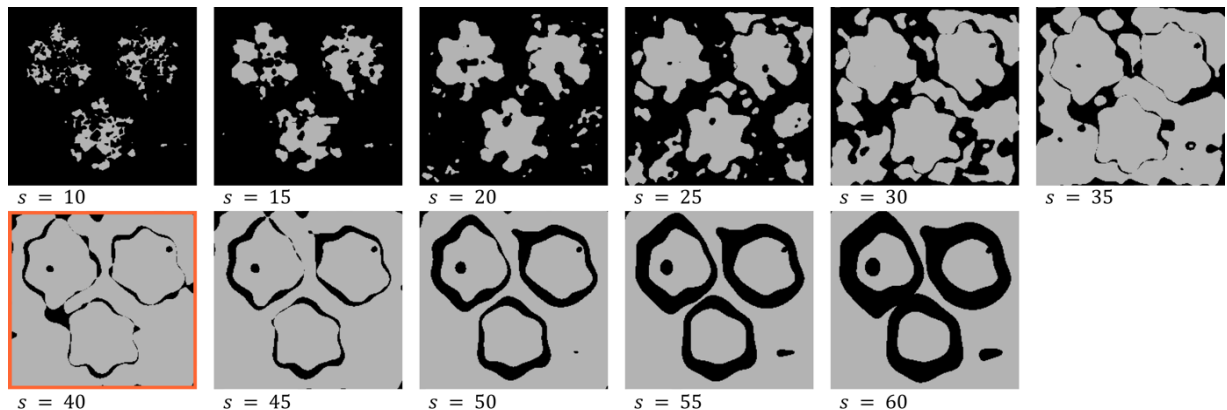

**Figure 6.** Example of a coverage computation for an instance of the bivariate snowflake model with noise $\gamma = 0.1$ and $N = 1600$. For each value $s$ of the smoothing parameter, the combined significant areas for all four spatial conditions $(Z_1, Z_2) \in \{(0,0), (1,0), (0,1), (1,1)\}$ as determined by a separate run of GeoSPM are shaded in light grey. The maximum number of significant grid cells is obtained for $s = 40$, highlighted in red.

The binary maps for each variable were assessed relative to their respective target maps, which were derived by thresholding the corresponding marginal distribution of the model, adding grid cells with a probability greater than 0.5 to the target. We applied a number of representative image segmentation metrics to each pair of maps, computing a mean score over the 10 repetitions of each unique triplet (model, $N$, $\gamma$) and variable. The following metrics were used (Dubuisson and Jain, 1994; Taha and Hanbury, 2015): Jaccard index, Dice score, Matthew's correlation coefficient, symmetric uncertainty and the modified Hausdorff distance (as a fraction of the length of the model diagonal). The mean score and deviation for each metric and computation method were aggregated into the plots reported below.

**2.4.3 Numerical Experiments: Interaction Parameterisation**

These experiments used the snowflake interaction model above and comprise observations of its variables, augmented by an interaction term. The interaction term is formed in the usual manner, by multiplying the observed values for both variables, yielding augmented observations: $\boldsymbol{y}' = (y_1, y_2, y_1 \cdot y_2)^\mathrm{T}$. The regional arrangement of the model is the same as the one employed for the bivariate snowflake model shown on the left of Figure 1. A single sampling level $N_{interaction} = 15000$ was used and the interaction parameter $c_3$ was increased from 0.25 to 0.5 in steps of 0.05. For each level of $c_3$, $R = 10$ independent data sets were randomly generated. We set a single value for the smoothing parameter $s = 60$, which was the highest value for the noise experiments. As before, a one-tailed $t$-test at $P < 0.05$ FWE (voxel-level family-wise correction) determined areas of significance and the same set of image segmentation metrics was computed for the binary maps.

**2.4.4 UK Biobank Experiments**

Results for the UK Biobank data were obtained by a single invocation of GeoSPM for each of the four models listed in Table 6. We choose a smoothing value of 7 kilometres as the 95 percent iso-density diameter of the bivariate normal distribution. This represents 20 percent of the width and height of our Birmingham analysis area, and seemed appropriate for identifying large-scale local variation. This time, a *two*-tailed $t$-test at $P <$ 0.05 FWE (voxel-level family-wise correction) was used for thresholding the statistic maps. Analysis is restricted to areas where the combined smoothing density of all observations is at least 10 percent of its maximum.

# 3 Results

Our numerical experiments with a known generative model enabled us to measure performance against a known ground truth under circumstances varying in density of sampling and contamination with noise. It also permits robust evaluation of graded interaction effects. In total, 2160 independent simulations with synthetic data were performed for the univariate models, 1440 for the bivariate models and 60 for the interaction model. Summarising scores within the three sets of simulations, we derive performance curves for GeoSPM and kriging solutions in each case. We then proceed to illustrate the application of GeoSPM to real world data from UK Biobank.

## 3.1 Synthetic models

Displayed in the following figures are sets of independent simulations comparing the performance of GeoSPM (in yellow) vs kriging (in green) as a function of contaminating noise, measured by five different indices of retrieval fidelity, using the snowflake (Figure 7) or anti-snowflake (Figure 8) bivariate ground truths, and low or high data sampling regimes (similar results for the univariate ground truths are reported in supplementary materials). A visual summary of the recovered binary maps underlying these performance curves—for the bivariate snowflake model and the high sampling regime—affords a further qualitative comparison between the two methods (Figure 9).

It is evident that GeoSPM offers superior efficiency across most of the noise range in all models and on all metrics. GeoSPM models generally remain stable at higher levels of noise than kriging. Both GeoSPM and kriging exhibit sensitivity to the sampling regime, both in terms of variability and stability but the effects are dwarfed by the difference between the two approaches. The type of ground truth has negligible impact. In addition, neither changing the (cross-) covariance function used for kriging from a Matérn function to a Gaussian nor applying a different regime for dealing with coincident observations—such as averaging—yields a discernible improvement to the performance of kriging in this context (see Figure 21 in supplementary materials).

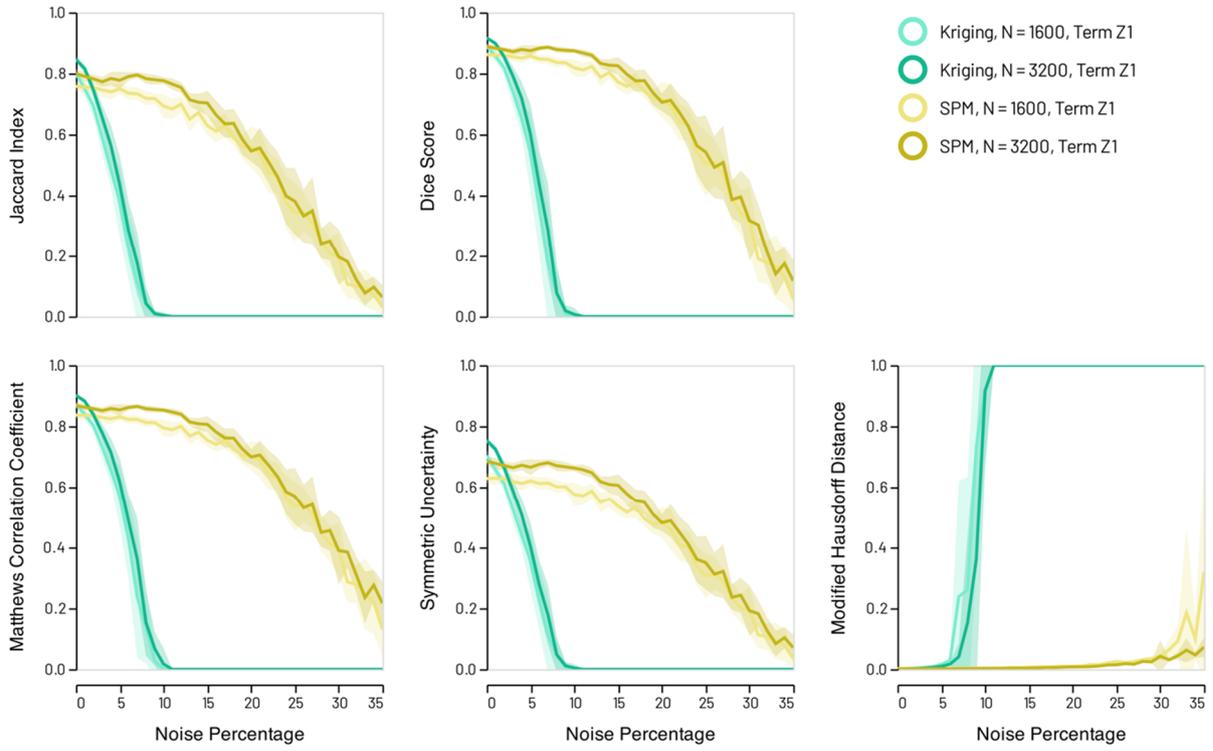

**Figure 7**. Synthetic snowflake models: Recovery scores for GeoSPM and kriging of model term $Z_1$ in the low (N = 1600) and high (N = 3200) sampling regimes. Lines denote the mean score across 10 random model realisations, shaded areas its standard deviation to either side of the mean. GeoSPM degrades more slowly and gracefully as noise increases compared with kriging.

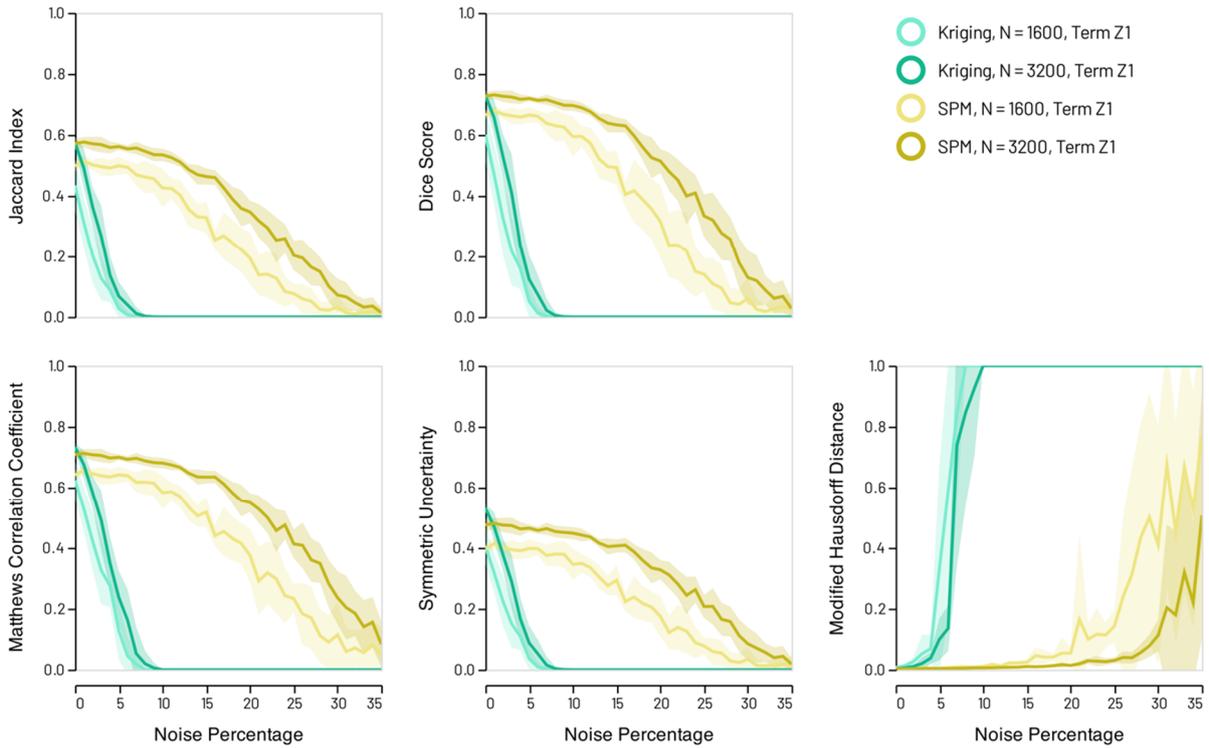

**Figure 8.** Synthetic anti-snowflake models: Recovery scores for GeoSPM and kriging of model term $Z_1$ in the low (N = 1600) and high (N = 3200) sampling regime. Lines denote the mean score across 10 random model realisations, shaded areas its standard deviation to either side of the mean. As is the case with the snowflake models, GeoSPM degrades more slowly and gracefully as noise increases compared with kriging.

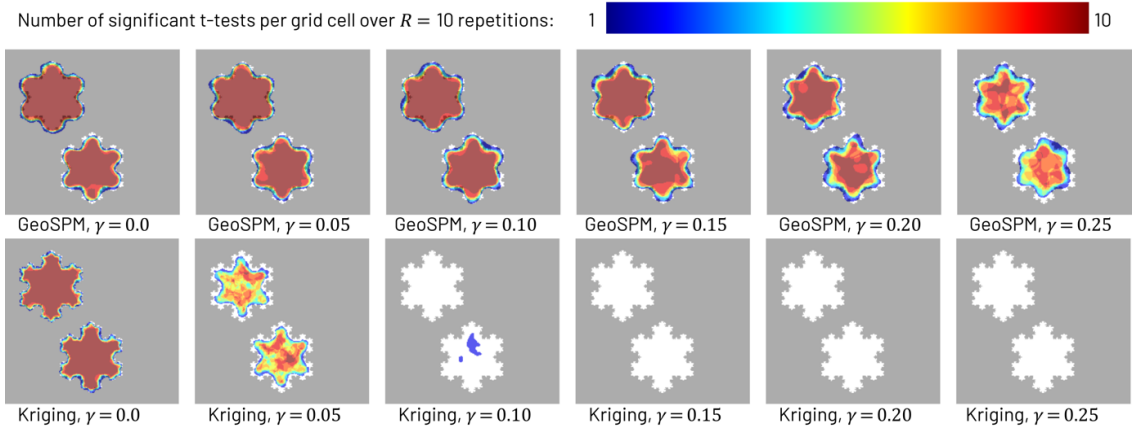

**Figure 9.** Recoveries of variable $Z_1$ in the synthetic bivariate snowflake model across $R$ = 10 repetitions for GeoSPM in the top row and kriging with a Matérn kernel and nugget component in the bottom row, both in the high sampling regime ($N$ = 3200). Grid cells that lie in the target region are shown in white, those outside in grey. The number of significant tests out of 10 repetitions is superimposed in colour for each grid cell: Dark blue indicates at least one significant test and dark red indicates the maximum number of 10, while cells with no significant test did not receive any colour. Kriging only produces recoveries up to a $\gamma$ value of 0.10, whereas GeoSPM still produces recoveries for much higher values of $\gamma$. GeoSPM used t-tests with a family-wise error corrected p-value of 0.05, for kriging we applied a z-test with an uncorrected p-value of 0.05, a null mean of 0.5 and a sample deviation obtained from the (positional) kriging variance estimate, as described in the section on "Synthetic Experiments: Noise Parameterisation".

The recoveries obtained from simulations of the interaction model clearly show GeoSPM's ability to detect an interaction between two spatially distributed factors, even towards the lower end of the approximate interaction effect size range (Figure 10). Plots of the same five indices above demonstrate successful retrieval for these interaction simulations quantitively (Figure 11). As we increase the size of the approximate interaction effect $c_3$, retrieval results for the interaction term $Z_1 \times Z_2$ approach those of the previous, noise-free bivariate snowflake model (setting aside the different sampling regimes). At the same time, recovery for variable $Z_1$ decreases in the interaction region $R_3$ (but not elsewhere), as the interaction term explains more variance. Once the recovery for variable $Z_1$ in region $R_3$ has vanished, the corresponding retrieval scores are about half of those for the same term in the noise-free model, which agrees with our expectation, because only one of two snowflake shapes in the target are still retrieved at that stage.

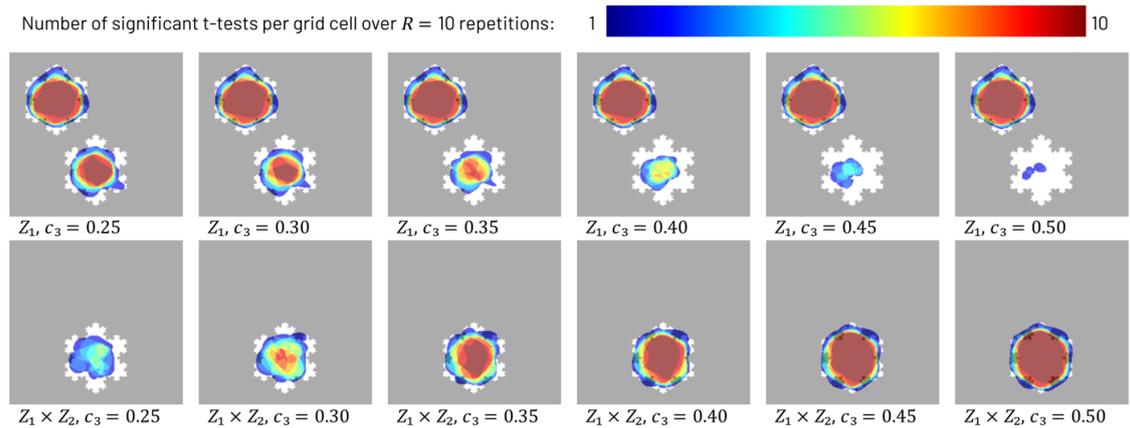

**Figure 10**. Recoveries produced by GeoSPM for the synthetic interaction model across $R$ = 10 repetitions for variable $Z_1$ in the top row and term $Z_1 \times Z_2$ in the bottom row, with $N$ = 15000 samples. Grid cells that lie in the target region are shown in white, those outside in grey. The number of significant tests out of 10 repetitions is superimposed in colour for each grid cell: Dark blue indicates at least one significant test and dark red indicates the maximum number of 10, while cells with no significant test did not receive any colour. Starting with a low value for the interaction effect $c_3$ on the left, recovery of the interaction term $Z_1 \times Z_2$ in region $R_3$ is weak, while recovery for variable $Z_1$ in the same region is stronger. This correlates with the fact that observations $(1,1)$ occur with only a slightly elevated probability $p_3 = 0.6$ compared with their null probability of 0.525 when $c_3$ equals 0 in the same setting. As $c_3$ increases towards the right, recovery in the same region for term $Z_1 \times Z_2$ increases ($p_3 = 0.725$ at the right), while recovery for variable $Z_1$ decreases (probability $p_1 = 0.125$ at the right for observing $(1,0)$, which is half of what it would be if there was no interaction effect). GeoSPM used t-tests with a family-wise error corrected p-value of 0.05.

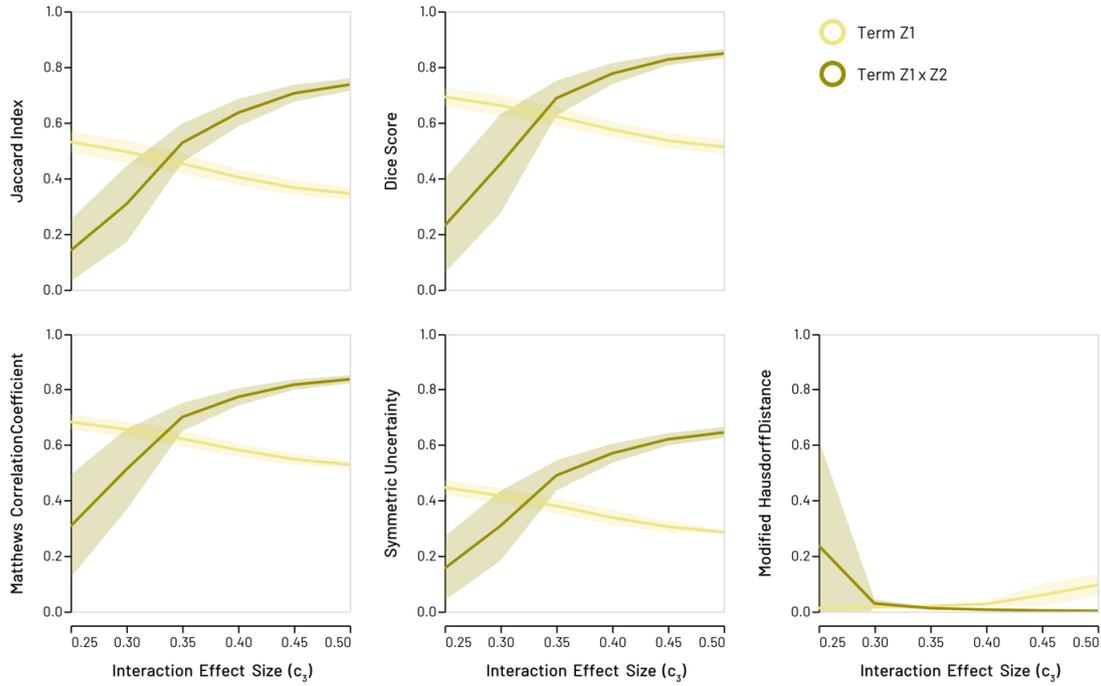

**Figure 11.** Synthetic snowflake interaction model: Recovery scores for SPM model variable $Z_1$ and term $Z_1 \times Z_2$ with N = 15000 samples. Lines denote the mean score across 10 random model realisations, shaded areas its standard deviation to either side of the mean. We increase the approximate interaction effect in region $R_3$ of the grid from left to right, so that the probability of observing $(1,1)$ grows while the probability of observing $(1,0)$ or $(0,1)$ shrinks (the probability of observing $(0,0)$ stays the same). As a result, scores increase for the interaction term $Z_1 \times Z_2$ as it captures more of the overall variance, whereas scores for variable $Z_1$ decrease, until the only significant recovery occurs in region $R_1$, which represents half of the target for $Z_1$ and explains why the overall decrease saturates.

### 3.2 UK Biobank models

In real-world scenarios there is usually no explicit ground truth against which an inference can be tested: the conclusion rests on the integrity of the underlying statistical assumptions. Our illustrative analysis of UK Biobank data therefore does not seek to quantify GeoSPM's fidelity but to demonstrate its potential utility in the medical realm. We focus on two aspects: the derivation of marginalised spatial maps that disentangle a factor of interest from a set of (interacting) confounders, and the use of conjunction analysis to identify regions jointly modulated by multiple spatially organised factors.

The propensity to develop type 2 diabetes is related to age, sex, BMI, and household income, amongst other factors: a known pattern clearly replicated in UK Biobank. A map of diabetes may therefore reflect not just the propensity to develop the disease but also the spatial structure of associated factors, both causal and incidental. If we are pursuing a previously unknown spatial factor—pollution, for example—we would wish to void our diabetes map of known confounders, yielding a spatial distribution of fully marginalised propensity.

We demonstrate GeoSPM on UK Biobank data drawn from Birmingham. Figure 12 presents the regression coefficient maps and significant *t*-test areas for four separate models of diabetes with incrementally greater numbers of covariates. The first, univariate, model of diabetes reveals an extensive concentric organisation, positive in the centre and negative in the periphery, especially in the north and south. The map becomes more tightly circumscribed with the addition of sex, age and BMI in model 2: and the two negative areas in the north and south are no longer significant, and a stronger negative region emerges west of the centre. With the addition of further covariates and their interactions, the spatial structure of diabetes converges on a set of focal, central locations, displayed in detail in comparison with the univariate model in Figure 13.

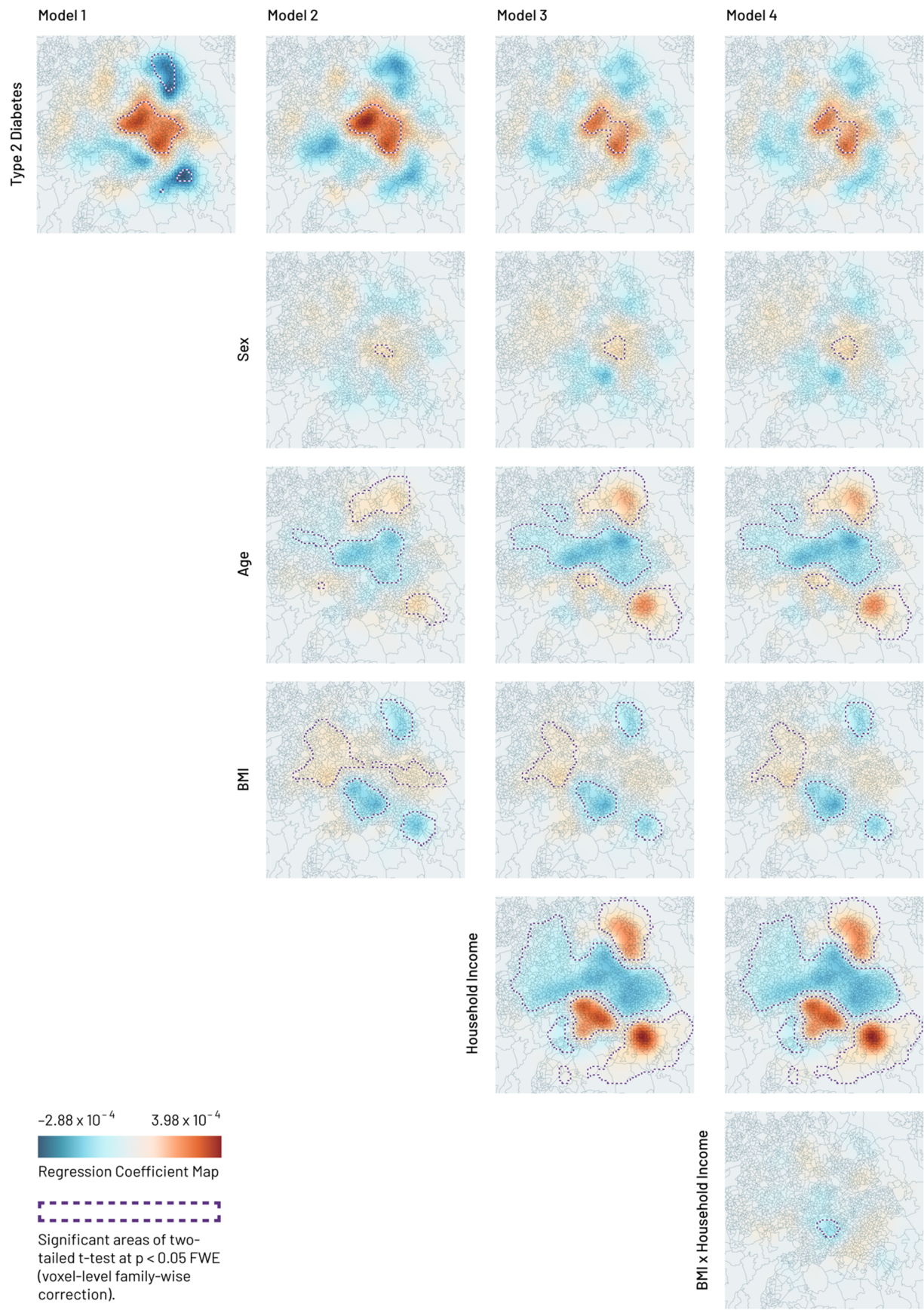

**Figure 12.** GeoSPM results for the four UK Biobank models of Birmingham (one column per model): Geographic regression coefficient maps are shown with outlines of significant areas in the corresponding two-tailed t-test at p < 0.05 FWE (voxel-level family-wise correction). The smoothing parameter value is 7000 meters.

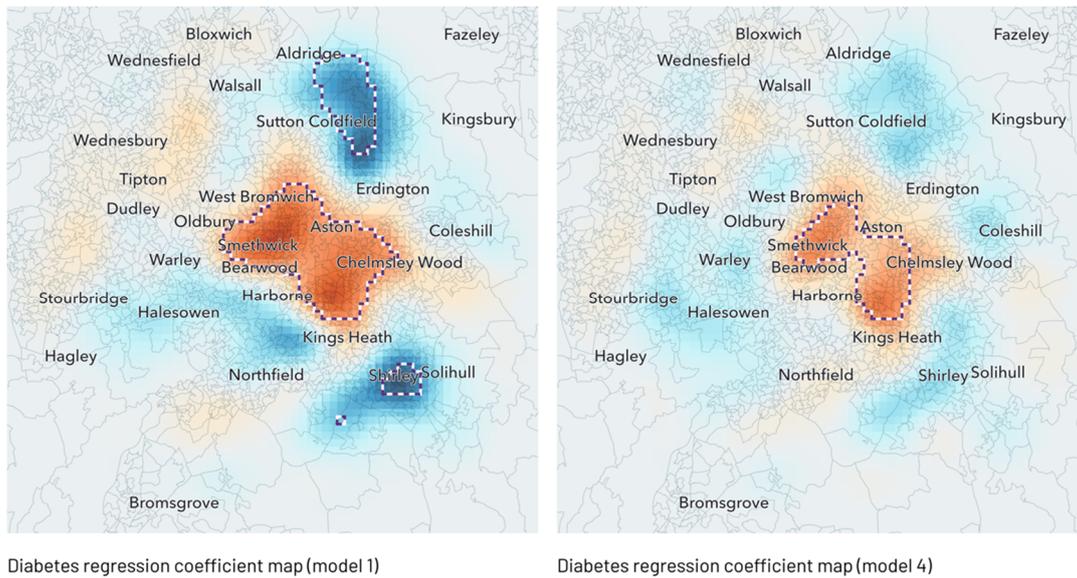

**Figure 13.** Geographic regression coefficient maps with location names for a single run of UK Biobank models 1 and 4: Model 1 is a univariate model of diabetes, model 4 adds sex, age, BMI, household income and an interaction term BMI x household income. Outlines show significant areas in the corresponding two-tailed t-test at p < 0.05 FWE (voxel-level family-wise correction). The smoothing parameter value is 7000 meters. The colour map scale is the same as in Figure 12 before.

We can now examine the *conjunctions* of multiple maps, not necessarily derived from the same model, within a second-level analysis. Conjunctions are here simply the intersections of two or more thresholded t maps, identifying areas where the regression coefficients and their associated variables are jointly significant. Applied to the outputs of our most complex model above, the approach and resulting conjunctions are shown in Figure 14. Pairwise conjunctions show a single region where diabetes and male sex are colocalised; two distinct adjacent regions where diabetes and age are inversely associated; and an identical pair of regions where diabetes is inversely related to household income. Finally, a three-way conjunction identifies a region populated by younger males of lower income associated with diabetes (Figure 15).

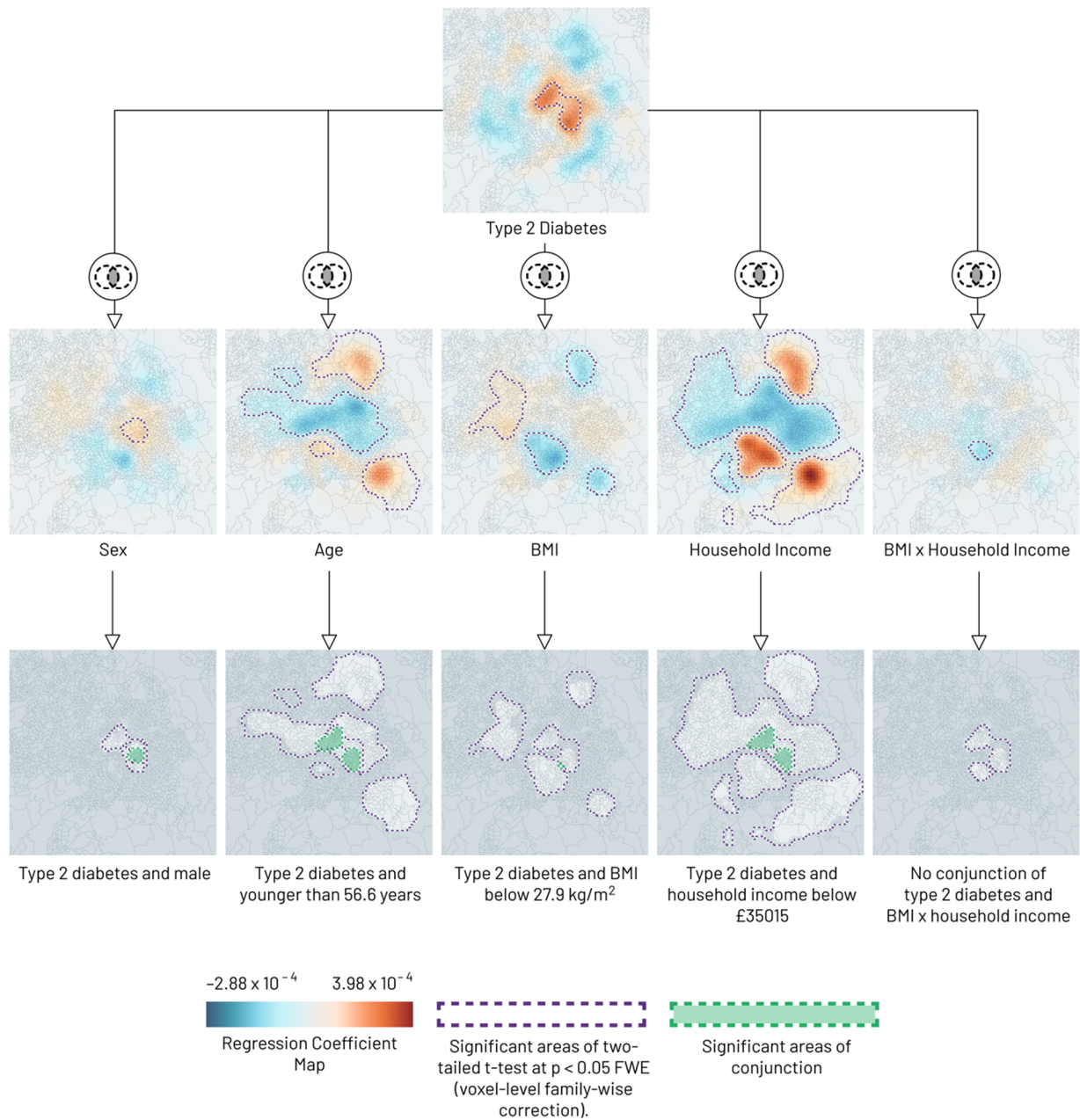

**Figure 14.** Binary conjunctions of geographic regression significance maps for a single run of UK Biobank model 4: A binary conjunction is formed of the significant areas of a two-tailed t-test at p < 0.05 FWE (voxel-level family-wise correction) between type 2 diabetes and, in turn, sex, age, BMI, household income and BMI × household income. Purple outlines show significant areas in the two-tailed t-test of each variable, green outlines show significant areas of conjunction: Significant areas of conjunction arise in diabetes combined with each of sex (male), age (younger than 56.6 years), BMI (below 27.9 kg/m²) and household income (below £35015). No significant areas of conjunction exist for diabetes and BMI × household income. Locations shown in darker grey tone are not significant for any of the variables. The smoothing parameter value is 7000 meters.

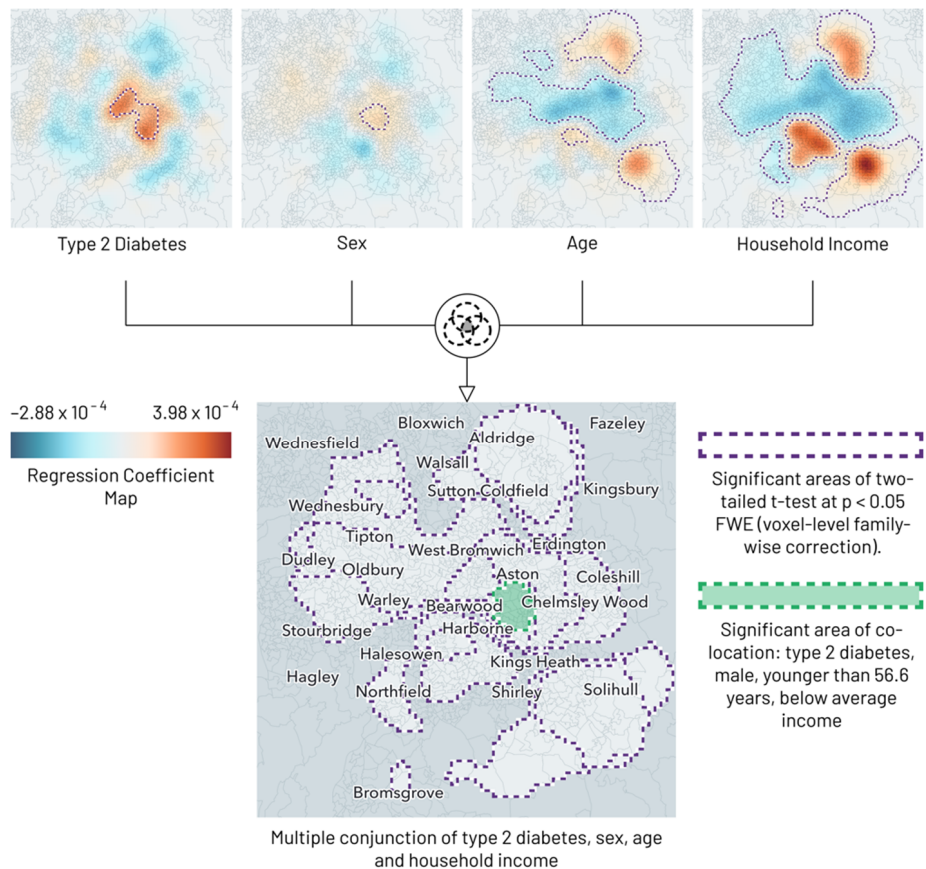

**Figure 15.** Example of a multiple conjunction (here quaternary) of geographic regression significance maps for a single run of UK Biobank model 4: A binary conjunction is formed of the significant areas of a two-tailed t-test at p < 0.05 FWE (voxel-level family-wise correction) between type 2 diabetes and, in turn, sex, age, BMI, household income and BMI × household income. Purple outlines show significant areas in the two-tailed t-test of each variable, green outlines show significant areas of conjunction: We can identify a significant area where younger males of lower income are associated with having type 2 diabetes in Birmingham. The smoothing parameter value is 7000 meters.

This concludes our illustration of GeoSPM. Note that the fact that GeoSPM was able to identify significant regionally specific effects provides a provisional form of predictive validity; under the assumption that these effects were present in the population—and could therefore be used to predict response variables.

## 4 Discussion

We have proposed, implemented, and validated a novel approach to drawing spatial inferences from sparse clinical data, extending to geostatistics a mature, principled inferential framework—statistical parametric mapping—that is well-established in the realm of brain imaging. Compared with the spatial models in current use, GeoSPM combines similar fidelity under optimal conditions with substantially less sensitivity to noise and under-sampling, greater robustness to failure, faster computation, graceful handling of multiple scales of spatial variation, and formal inferential support. Crucially, its simplicity and accessibility facilitate widespread application of the comprehensive software implementation we have provided, built on software validated SPM open-source codebase, across a wide range of applications in medicine and beyond. Here we consider six points concerning the application, extension, and limitations of our approach.

First, GeoSPM is applicable to any problem of spatial inference, whose formulation conforms to the minimal assumptions of the underlying statistical framework. The types of data, the choice of model evaluated at each point, and the size and density of the evaluated grid are not under any strong constraint. Eliminating the spatial dimension allows each point-wise model to be more flexible than the data or computational resource could otherwise sustain. The model could even be complicated spatially, extending to encompass a local patch within otherwise the same framework. This is a key strength in medical applications, where a spatial effect typically needs to be disentangled from a wide array of others.

Second, though here prototyped on stationary data, GeoSPM can be configured with time instead of the spatial scale in the third dimension, enabling graceful modelling of both spatial and temporal correlations. This has been used, for example, in the context of electrophysiology (Litvak *et al.*, 2011) where extra dimensions can include peristimulus time or, indeed, fast oscillatory frequencies. The effects of manipulating noise and spatial dependencies can then be evaluated across individual time series. Equally, the third dimension could be used for multimodal data projected within the same grid, informing the inference by multiple sampling modalities.

Third, the smoothing parameter may be constrained by prior knowledge or independent estimation from the data, even if evaluating a set of models over a plausible range is arguably the most robust approach. One may alternatively rely on the properties of the inferred maps, as suggested in our validation analyses. All competing spatial modelling frameworks rely on chosen parameters to some degree; ours is reduced to a single readily interpretable one.

Fourth, no model could perfectly remedy defects in the data itself, such as inadequate or biased coverage. The former can be mitigated by confining inference to spatial locations exhibiting sufficient sampling density; the latter, analogously to structured missingness, is not easily remediable within this or any other inferential framework, and presents no more or less of a problem.

Fifth, GeoSPM, like SPM itself, is a platform for standard frequentist statistical inference, revealing the organisation of spatially structured variables without causal implications of any kind. But, also like SPM, it is open both to Bayesian extensions, and causal modelling downstream of the core framework. There are many ways of querying data, both with classical mass univariate and Bayesian analyses of this kind. Although not illustrated here, model comparison using the *F*-statistic is a common application that could be enabled by GeoSPM. For example, one could ask whether household income has an effect on the regional prevalence of diabetes, having accounted for other demographic variables, by comparing (general linear) models that do and do not include household income as an explanatory variable.

Finally, the SPM approach, in any formulation, is designed for topological inference, not discrimination between distributed spatial patterns, which requires explicit modelling of spatial interactions that only a multivariate model could conceivably deliver. Indeed, such use would violate the underlying assumption of benign regional dependence, as do analogous attempts in the domain of lesion-deficit mapping of the brain(Mah *et al.*, 2014). GeoSPM maps may nonetheless be used to select features where the fragility of the multivariate model, or the applicable data regime, compel it.

## Acknowledgements

This work is aligned with a project on 'Novel methods to explore the value of cognitive health in a place' supported by the Health Foundation, an independent charity committed to bringing about better health and health care for people in the UK. Holger Engleitner, Amy Nelson, Daniel Herron, and Parashkev Nachev are supported by the NIHR UCLH Biomedical Research Centre. Marta Suarez Pinilla is supported by the Health

Foundation. Martin Rossor is supported by the National Institute for Health Research (NIHR) Senior Investigator Award/NF-SI-0512-10033. Ashwani Jha, Geraint Rees, and Parashkev Nachev are supported by Wellcome (213038). The views expressed are those of the authors and not necessarily those of the NIHR, the Department of Health and Social Care, the Health Foundation, or Wellcome. The funders had no role in study design, data collection and analysis, decision to publish, or preparation of the manuscript.

# Supplementary Materials

## S.1 Bibliographic analysis

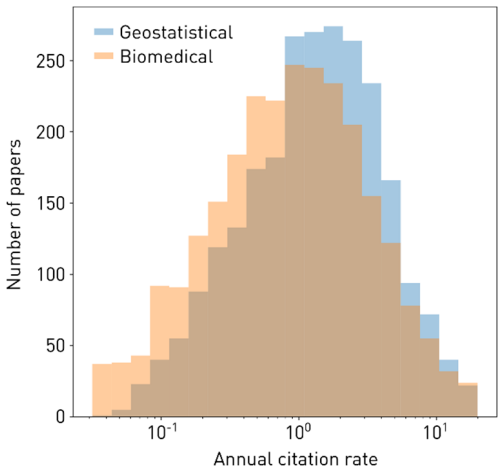

**Figure S1.** Overlapped histograms of the decimal log-transformed annual citation rates of 1897 identified journal papers at the intersection of spatial analysis and medicine cited more than once (blue), and an identically filtered random sample of non-spatial biomedical papers (orange), published between 1990 and 2019. The untransformed distributions are significantly different on a Mann-Whitney U test, p<0.001.

## S.2 Additional Results

### S.2.1 Synthetic Experiment Results for Term $Z_2$ of bivariate models

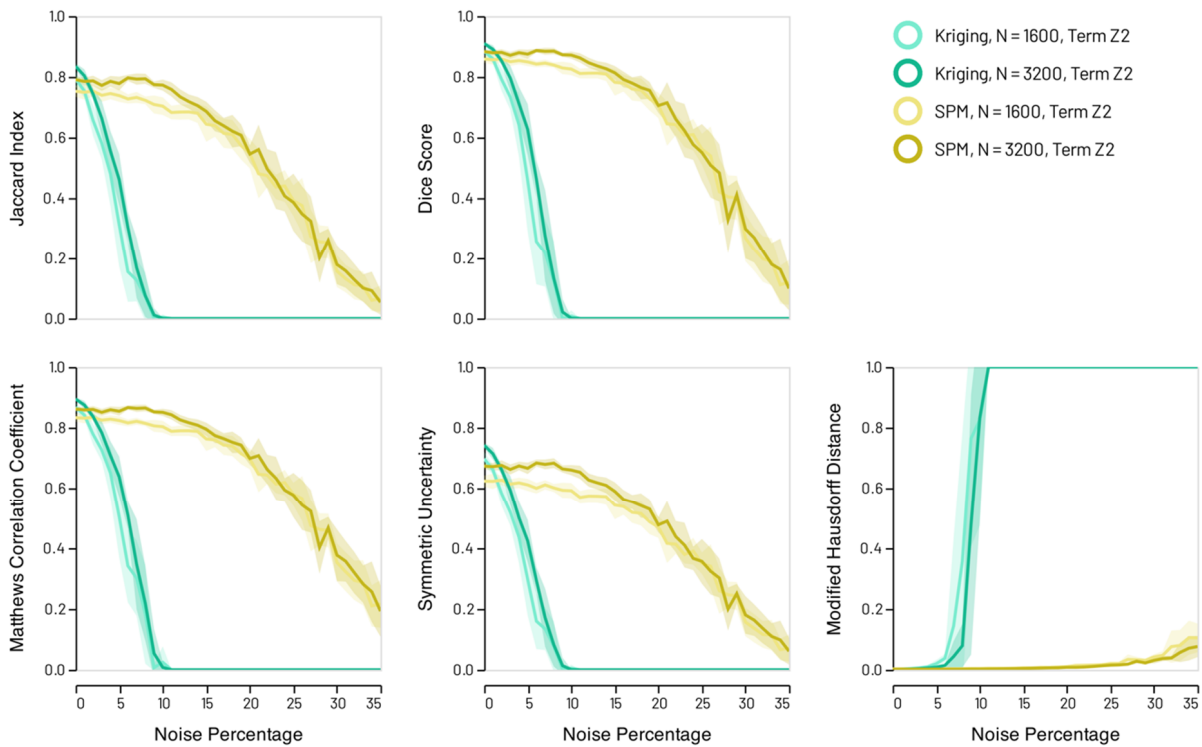

**Figure S2.** Synthetic snowflake models: Recovery scores for GeoSPM and kriging model term $Z_2$ in the low (N = 1600) and high (N = 3200) sampling regime. Lines denote the mean score across 10 random model realisations, shaded areas its standard deviation to either side of the mean. GeoSPM degrades more slowly and gracefully as noise increases compared to kriging.

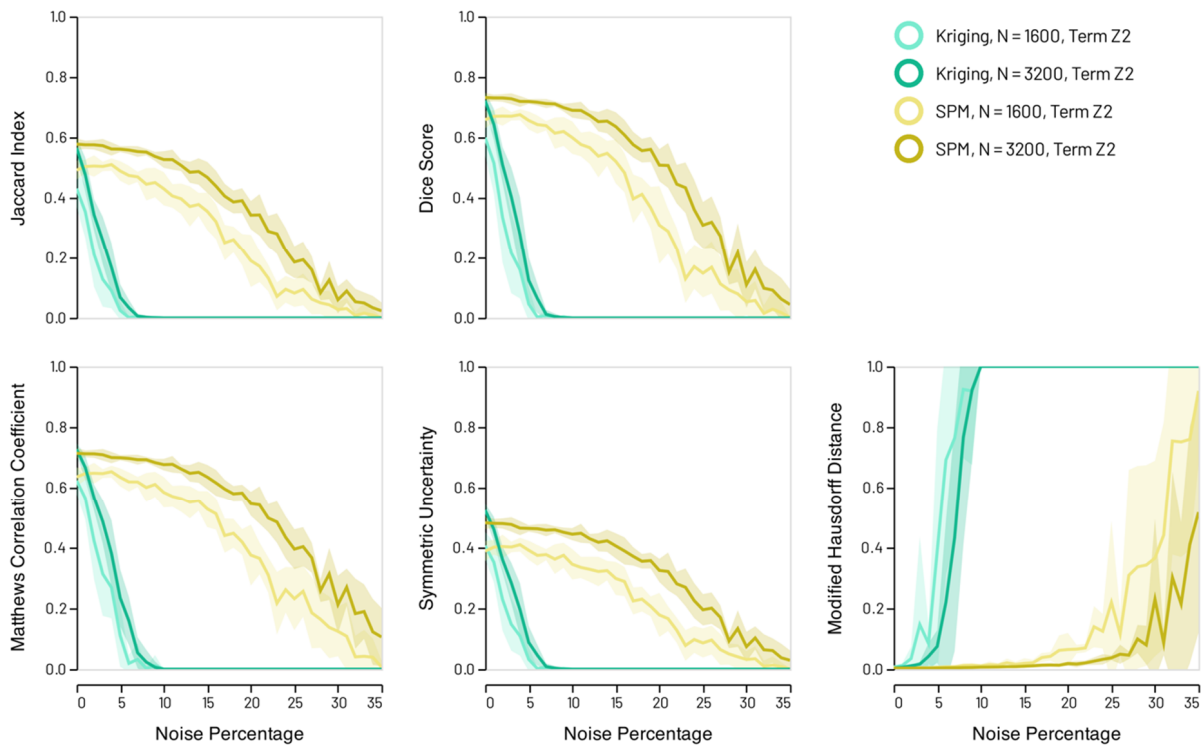

**Figure S3.** Synthetic anti-snowflake models: Recovery scores for GeoSPM and kriging model term $Z_2$ in the low (N = 1600) and high (N = 3200) sampling regime. Lines denote the mean score across 10 random model realisations, shaded areas its standard deviation to either side of the mean. GeoSPM degrades more slowly and gracefully as noise increases compared with kriging.

### 6.1.2 Synthetic Experiment Results for univariate models

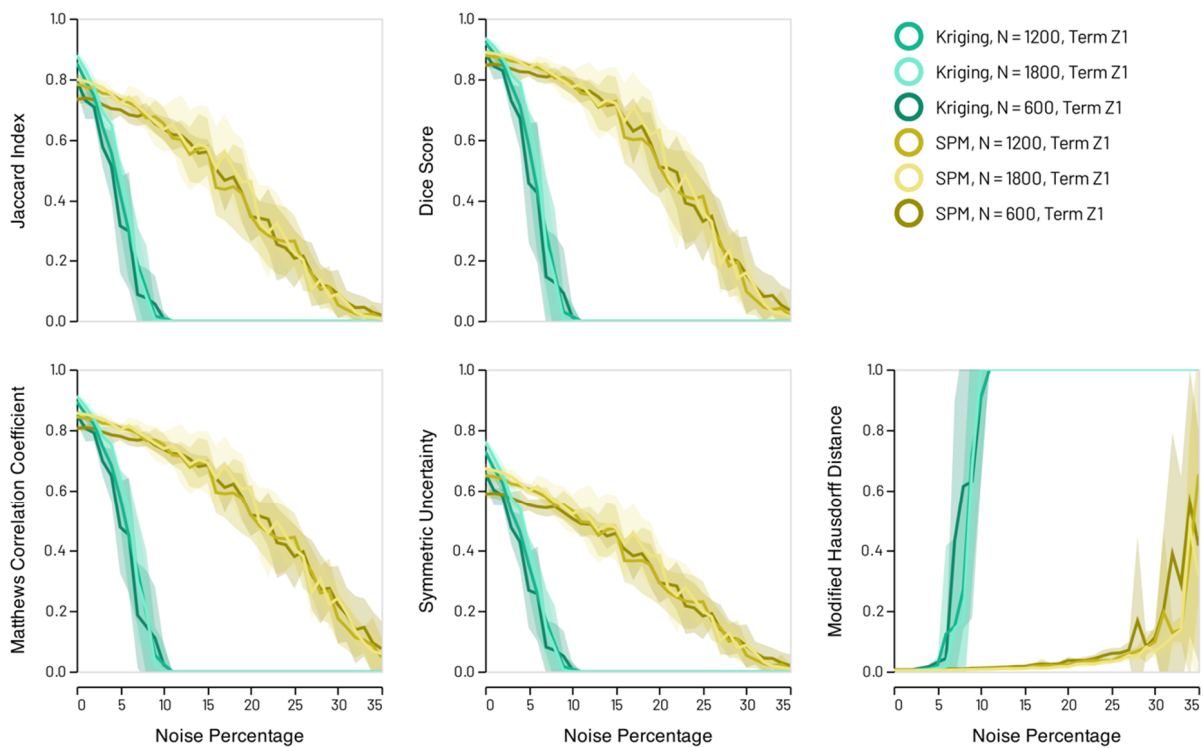

**Figure S4.** Synthetic univariate snowflake models: Recovery scores for the single GeoSPM and kriging model term in the low (N = 600), middle (N = 1200) and high (N = 1800) sampling regime. Lines denote the mean score across 10 random model realisations, shaded areas its standard deviation to either side of the mean. GeoSPM degrades more slowly and gracefully as noise increases compared with kriging.

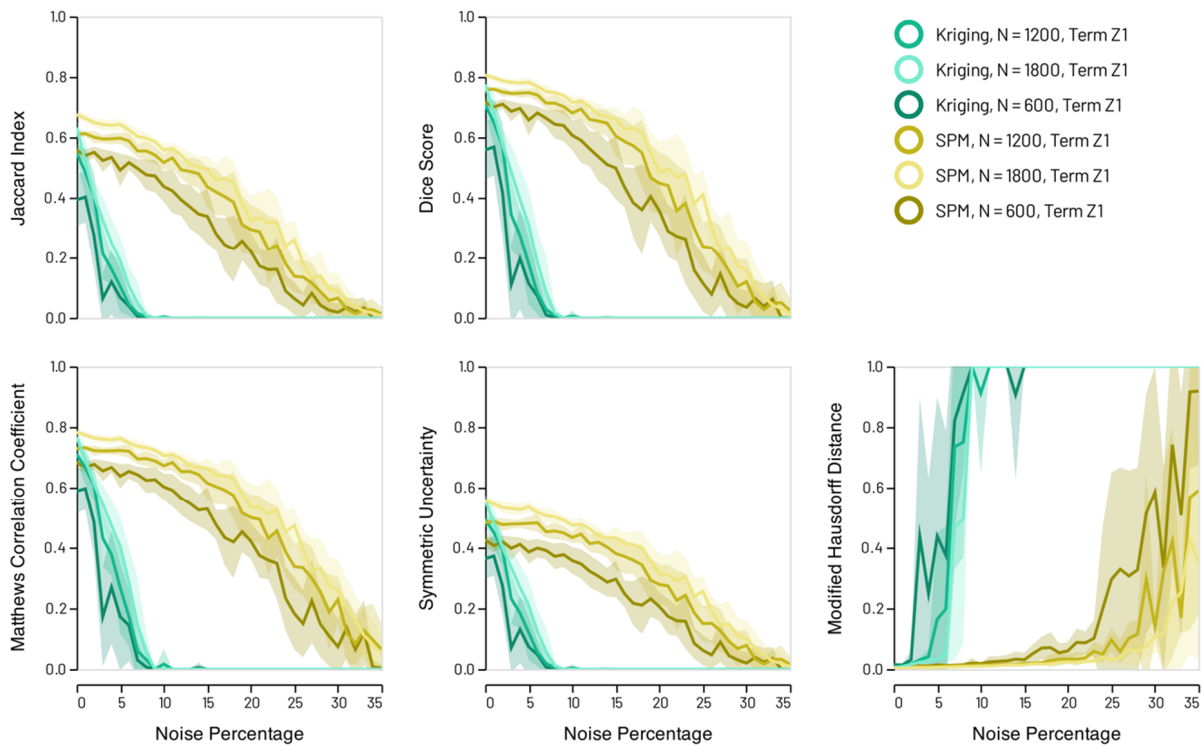

**Figure S5.** Synthetic univariate anti-snowflake models: Recovery scores for the single GeoSPM and kriging model term in the low (N = 600), middle (N = 1200) and high (N = 1800) sampling regime. Lines denote the mean score across 10 random model realisations, shaded areas its standard deviation to either side of the mean. GeoSPM degrades more slowly and gracefully as noise increases compared with kriging.

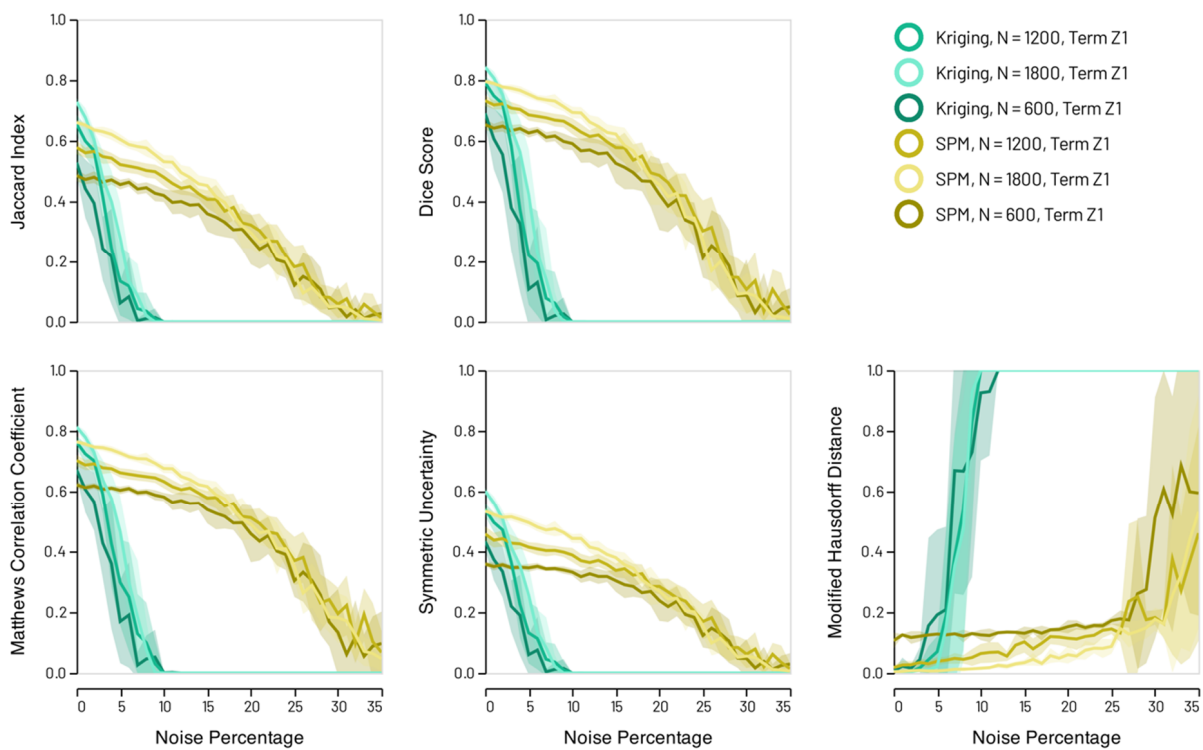

**Figure S6.** Synthetic univariate snowflake field models: Recovery scores for the single SPM and kriging model term in the low (N = 600), middle (N = 1200) and high (N = 1800) sampling regime. Lines denote the mean score across 10 random model realisations, shaded areas its standard deviation to either side of the mean. GeoSPM degrades more slowly and gracefully as noise increases compared with kriging.

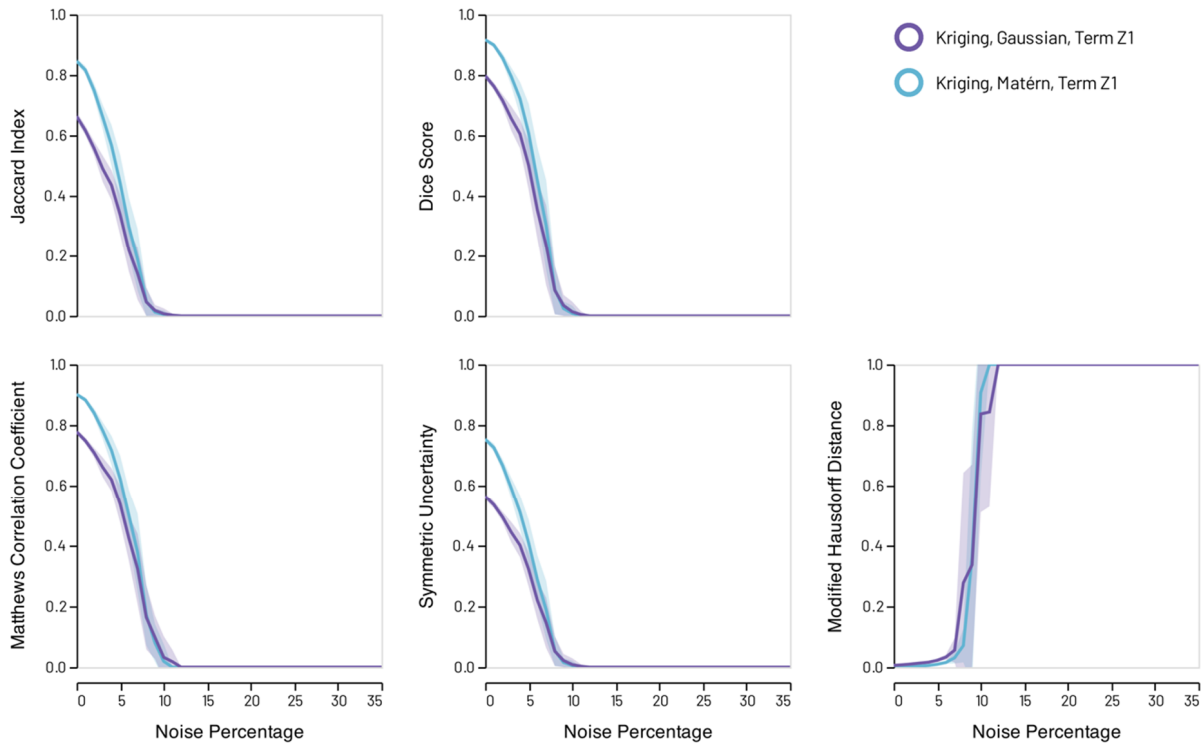

**Figure S7.** Synthetic bivariate snowflake models: Recovery scores for kriging model term $Z_1$ with a Matérn covariance function (blue) and a Gaussian covariance function (purple) in the high (N = 3200) sampling regime. Lines denote the mean score across 10 random model realisations, shaded areas its standard deviation to either side of the mean. In both cases coincident observations were *averaged* and reduced to one instead of adding a small amount of random noise to their locations as before. However, this did not change the performance in any meaningful way when compared with a Matérn covariance function with random noise added (as shown in Figure 7): That curve is almost identical to the averaged version displayed here in blue and was therefore left out. The Gaussian covariance function performs slightly worse than the Matérn covariance. This leads us to believe that kriging performance is not improved in our experiments by choosing a different covariance function or coincident observation regime.

## ICD-9 codes

| ICD-9 Group | Description | ICD-9 Codes in Group |
|---|---|---|
| 250.0 | Diabetes mellitus without mention of complication | 250.00, 250.02 |
| 250.1 | Diabetes with ketoacidosis | 250.10, 250.12 |
| 250.2 | Diabetes with hyperosmolarity | 250.20, 250.22 |
| 250.3 | Diabetes with other coma | 250.30, 250.32 |
| 250.4 | Diabetes with renal manifestations | 250.40, 250.42 |
| 250.5 | Diabetes with ophthalmic manifestations | 250.50, 250.52 |
| 250.6 | Diabetes with neurological manifestations | 250.60, 250.62 |
| 250.7 | Diabetes with peripheral circulatory disorders | 250.70, 250.72 |
| 250.8 | Diabetes with other specified manifestations | 250.80, 250.82 |
| 250.9 | Diabetes with unspecified complication | 250.90, 250.92 |

**Table S1.** ICD-9 codes used in the extraction of the type II diabetes indicator variable.